\documentclass[journal]{IEEEtran}

\ifCLASSINFOpdf
   \usepackage[pdftex]{graphicx}
   \DeclareGraphicsExtensions{.pdf,.jpeg,.png}
\else
  \usepackage[dvips]{graphicx}
  \DeclareGraphicsExtensions{.eps}
\fi

\graphicspath{{./}}
\usepackage[ruled,vlined,linesnumbered]{algorithm2e}
\usepackage{epstopdf}
\newcommand{\figref}[1]{\figurename~\ref{#1}}
\newcommand\norm[1]{\left\lVert#1\right\rVert}
\usepackage{amsmath}
\usepackage{array}
\usepackage{xspace}
\usepackage{url}
\usepackage[english]{babel}
\usepackage[utf8x]{inputenc}
\usepackage{amsmath}
\usepackage{graphicx}
\usepackage[colorinlistoftodos]{todonotes}
\usepackage{amssymb}
\usepackage{amsfonts}%
\usepackage{graphicx}
\usepackage{multirow}
\usepackage{epsfig}
\usepackage{amsthm}
\usepackage{graphicx}
\usepackage{psfrag}
\usepackage{mcode}
\usepackage{tabularx}
\DeclareMathAlphabet\mathbfcal{OMS}{cmsy}{b}{n}
\newcommand*\Bell{\ensuremath{\boldsymbol a}}
\newcommand*\Br{\ensuremath{\boldsymbol b}}
\newtheorem{thm}{Theorem} 
\theoremstyle{thm}
\newtheorem{defn}{Definition}  
\theoremstyle{defn}
 
\theoremstyle{exmp}
\newtheorem{lem}{Lemma} 
\theoremstyle{lem}
\newtheorem{rem}{Remark} 
\theoremstyle{rem}
 
\theoremstyle{coro}
 
\theoremstyle{prop}
\begin{document}
\title{QUARKS: Identification of large-scale Kronecker Vector-AutoRegressive models}
\author{Baptiste~Sinquin$^1$
        and~Michel~Verhaegen$^1$
\thanks{$^1$ Both authors are with the Delft Center for Systems and Control, Technische Universiteit Delft, the Netherlands.}
\thanks{Corresponding author: baptiste.sinquin@gmail.com}
\thanks{The research leading to these results has received funding from the European Research Council under the European Union's Seventh Framework Programme (FP7/2007-2013) / ERC grant agreement 339681.}}

\maketitle

\begin{abstract}
In this paper, we address the identification of two-dimensional spatial-temporal dynamical systems described by the Vector Auto-Regressive (VAR) form. The coefficient matrices of the VAR model are parametrized as sums of Kronecker products. When the number of terms in the sum is small compared to the size of the matrices, such a Kronecker representation efficiently models large-scale VAR models. Estimating the coefficient matrices in least-squares sense gives rise to a bilinear estimation problem which is tackled using an Alternating Least Squares (ALS) algorithm. 
Regularization or parameter constraints on the coefficient matrices allows to induce temporal network properties such as stability as well as spatial properties such as sparsity or Toeplitz structure.
Convergence of a particular formulation of ALS which features some normalization is proved using fixed-point theory.
A numerical example demonstrates the advantages of the new modeling paradigm. It leads to comparable variance of the prediction error with the unstructured least-squares estimation of VAR models. However, the number of parameters grows only linearly with respect to the number of nodes in the 2D sensor network instead of quadratically in the case of fully unstructured coefficient matrices.
\end{abstract}


\IEEEpeerreviewmaketitle

\section{Introduction}
System identification of large-scale networks has received an increased attention during the past years. Major scientific and engineering projects such as \cite{elt} challenge the identification community to derive scalable identification algorithms. In this work, the focus is on regular 2D sensor networks defined on a square or circular rastered grid of size $N \times N$ with N large. Such networks occur in multi-dimensional signal processing problems, such as image/video processing \cite{Derado:2010} but also in control for high resolution imaging using adaptive optics \cite{AO}. Prior knowledge on the network structure is often of prime importance to cope with the challenges inherent to the large dimensions. 
Potential model structures for representing the spatial-temporal dynamics include 2D state-space models, e.g  \cite{Roesser:75} \cite{Fornasini:78}, although the identification of the latter model in a system identification or convex optimization context is still an open research question. The difficulties that arise in e.g the Roesser model is that spatial causality is not present, hence making it a challenge to derive efficient identification methods and global convergence under generic conditions. For specific conditions, such as assuming a separability condition of the transfer function called Causal, Recursive, Separable in Denominator (CRSD), a subspace algorithm is derived in \cite{Mercere:2016}. The temporal dimension as well as considering spatial varying dynamics in a global (modal) context is not investigated however. The global context is of key relevance to impose constraints such as stability of the global network. Another framework consists in assuming that each node is connected to very few other nodes in the network. Identification of these \emph{sparse} networks has been investigated in \cite{Chiuso:2012} within the Bayesian framework.
\emph{Sequentially Semi-Separable} (SSS) networks are composed of a 1D string of interconnected systems and have been analyzed in \cite{Rice:2010}. Each of the subsystem is modeled with a mixed causal anti-causal linear time varying model and shares unknown interconnections with the closest neighbors. This framework reveals to be very elegant to deal in a scalable manner with large strings of subsystems: both linear algebra operations and control to achieve global $\mathcal{H}_2$ performance were shown to be achievable within linear computational complexity in the string's size. Subspace system identification for such systems has been achieved locally in \cite{Yu:16}.
The extension of the 1D SSS methods to higher spatial dimensions gives rise to multi-level SSS problems, for which up till now no efficient solution for identification (and control) exist. An alternative is proposed in \cite{Ali:2009} that introduces an identification method using 2D Finite Impulse Response models.
However, this identification and related methods search for a local model estimation and have difficulties in assuming and/or imposing global network properties such as the stability of the overall system. 
Rather than having a zonal representation of networks as in the previous paragraph, the work in \cite{Massioni:2008} sheds the light onto the modal subsystems associated with the network. 
A generalization is found in \cite{Mass:14} which introduces \emph{$\alpha$-decomposable} networks that allow for $\alpha$ different kinds of subsystems in the whole network to interact. In this paper, we present a modal network representation that includes $\alpha$-decomposable systems as a special case and allow the subsystems to be heterogeneous.

In this paper, a novel modeling and identification paradigm is introduced to model 2D spatial systems with temporal dynamics. As a fundament of this new approach, we restrict to temporal Vector Auto-Regressive (VAR) models with the spatial structure imposed on the coefficient matrices. Let a 2D network connected on a grid of size $N \times N$. The spatial structure represents the coefficient matrices $\{\bold{A}_i \}_{i = 1..p}$ as a finite sum of a Kronecker product between low dimensional matrices:
\[
\bold{A}_i = \sum_{i = 1}^r \bold{U}_{i} \otimes \bold{V}_{i} \quad \in \mathbb{R}^{N^2 \times N^2}
\]
where $r \ll N$ is called the Kronecker rank and $\bold{U}_i, \bold{V}_i \in \mathbb{R}^{N \times N}$ are the factor matrices. Such representation of large dimension matrices was studied in \cite{Loan:92} in which the equivalence between expressing a matrix as a sum containing few Kronecker products and a low-rank approximation of a reshuffled matrix was established. More than only enjoying the storage of a reduced number of entries, such a structure enables fast computations thanks to the very pleasant algebra of the Kronecker product, see e.g \cite{Loan:2000}. 

Using Kronecker structures for forward modeling of multi-dimensional problems is well-known, especially in tensor-based scientific computing \cite{bonn}. Besides, Kronecker structures have been applied efficiently for computing second moments in multi-dimensional processes \cite{Tsi:13}, for analyzing EEG signals \cite{Bijma:2005} and for image deblurring, \cite{Hansen:06}. The latter example enables to relate the Kronecker rank-one modeling with physical properties of the system. Denoting an object $\bold{O}$ imaged with a static optical system, the resulting blurred image $\bold{B}$ undergoes the linear blurring operation as follows: 
\begin{equation}
\label{psf}
\text{vec}(\bold{B}) = \bold{A}\text{vec}(\bold{O})
\end{equation}
The coefficient-matrix $\bold{A}$ is related to the Point-Spread Function (PSF) (or 2D impulse response) of the optical system. The equation \eqref{psf} represents the 2D convolution operation between the PSF and the object $\bold{O}$. The structure in $\bold{A}$ is related to the separability of the PSF, which implies the following Kronecker structure for the coefficient-matrix $\bold{A}$: 
\begin{equation}
\label{kron_intro}
\bold{A} = \bold{A_r} \otimes \bold{A_c}
\end{equation}
where $\bold{A_r}$ and $\bold{A_c}$ represent respectively the 1D convolution with the rows and columns. A large-scale static input-output map in \eqref{psf} is represented by a Kronecker matrix as in \eqref{kron_intro}. In a more general context, separation-of-variable techniques have been applied in \cite{doostan} and the references therein to break down the curse of dimensionality when modeling high-dimensional partial differential equations.

Although tensor-based algorithms for handling large datasets receive a growing interest, system identification of multi-dimensional systems is however in its infancy. 
An overview of data-driven algorithms that handle large datasets using the tensor representation was provided in \cite{tensors} among which a multilinear tensor regression for relational longitudinal data, \cite{hoff}. The approach proposed in \cite{hoff} handles the estimation of factor matrices from an input-output tensor model and using Alternating Least Squares. However, \cite{hoff} embeds temporal dynamics in a higher-order tensor whereas the parametrization we propose follows the control engineering approach to combine the temporal dynamics linearly while modeling independently each coefficient matrix with a sum of Kronecker matrices. Besides, we allow the Kronecker rank to be strictly larger than one for more generality and applicability for identification and control of systems such as adaptive optics. These two points are crucial to achieve good accuracy estimations in e.g a laboratory environment and hence, enable its effective use for control. Third, the QUARKS methodology proposes regularization to estimate stable and sparse models.

Another work related to the framework we propose deals with blind source separation using tensor representations, \cite{bss}. The approach consists in estimating two matrices $\bold{M}$ and $\bold{S}$ from the measurements stored in $\bold{X}$ given the relationship:
\begin{equation}
\label{bss}
\bold{X} = \bold{MS}
\end{equation}
where $\bold{M}$ represents the mixing matrix and $\bold{S} \in \mathbb{R}^{n \times K}$ the $n$ source signals for $K$ time samples. 
The work \cite{bss} relies on a low-rank decomposition of a certain reshaping (equivalently, segmentation) for either/both the rows of the mixing matrix and the source channels in order to achieve a trade-off between data compression and accuracy of the data fit. %The 2D spatial signals sampled on a regular grid that we analyze shares similarities with the segmentation framework in \cite{bss}. 
Both the present paper and \cite{bss} reshuffle the mixing vectors/coefficient matrices in order to exhibit a low-rank matrix and consequently, reduce the number of modeling parameters. %and establish a new trade-off between the accuracy of the representation and the data compression achieved. 
Nonetheless, our modeling assumptions differ in three ways. We model the coefficient-matrices with lower-dimensional matrices without making restrictive assumptions on the signals rather than being obtained from a regular grid and being persistently exciting. We focus on the specific case where the sources signals $\bold{S}$ are known which allows to get rid of the ambiguity transformation inherent to BSS identification and to formulate spatial and temporal stability constraints on the coefficient-matrices $\bold{A}_i$. Last, we exploit the 2D structure of the network and separability of the modeled functions in order to reduce the number of parameters. This point is detailed in Section III. The different modeling assumptions lead to distinct optimizations procedures.

In the following, the class of \emph{low-Kronecker rank} matrices is studied with a focus on modeling 2D spatial-temporal dynamical systems of Vector Auto-Regressive form.
The Kronecker tool as presented in this paper is meant to break down the curse of dimensionality when working with arrays of higher dimensions and without necessarily enforcing \emph{a priori} a sparsity pattern in the network, hence allowing to discover both spatially varying dynamics and an unknown topology from the data. It also serves as the basis for other more useful identification approaches such as subspace identification, see e.g \cite{hadron}. As such, it will establish the fundamentals of a new modeling framework for the identification and analysis of large-scale 2D dynamical systems. 
The challenge lies in deriving algorithms that are, on the one hand, scalable in terms of data storage as well as in terms of computational complexity in identifying and using these models, e.g in subsequent control design, and on the other hand, that still ensures similar prediction performances compared to the unstructured least-squares estimates.
The main contributions of this paper are the definition of a new class of dynamical systems -of low Kronecker rank-, the formulation of a regularized cost function for identification and the formulation of an Alternating Least Squares algorithm with $\mathcal{O}(N^3N_t)$ computational complexity where $N_t$ is the number of temporal samples.
\newline
The paper has the following outline. Section II describes the class of sums-of-Kronecker matrices, while Section III associates a VAR model associated with network data. 
In Section IV we describe regularization methods to emphasize the identification of stable models both in time and space. We study in Section V the Alternating Least Squares algorithm with a focus on the conditions to ensure global convergence. 
The methods are then illustrated in Section VI on a random low-Kronecker rank VARX model and a practical scenario dealing with open-loop identification of the atmospheric turbulence for adaptive-optics purposes.
\newline
\emph{Notations.}
Scalars are denoted by lower or uppercase letters or symbols. 
Vectors are written as boldface lower-case letters such as $\mathbf{x}$. The boldface is used to make a distinction between indexing a set of vectors, such as $\mathbf{x}_1, \mathbf{x}_2$, and referring to the elements of a single vector $\mathbf{x} \in \mathbb{R}^n$, such as $x_1, \hdots, x_n$. The null vector and the vector of ones is denoted by $\mathbf{0}$ and $\mathbf{1}$ respectively, where an index can be used to explicitly show its size e.g $\mathbf{1}_n \in \mathbb{R}^n$. The Euclidean norm of a vector $\mathbf{x}$ is written as $\| \mathbf{x} \|_2 = \sqrt{x_1^2 + \hdots + x_n^2} = \langle \mathbf{x},\mathbf{x} \rangle$. The sum in absolute value for the elements in $\bold{x} \in \mathbb{R}^n$ is denoted with $\| \bold{x}\|_1 = \sum_{i = 1}^n |x_i|$.
\newline
Matrices are represented by boldface uppercase letters such has $\mathbf{X}$. The element located at the $i$-th row and $j$-column of the matrix $\mathbf{X}$ is written as $x_{i,j}$, or $x_{\star,(i,j)}$ when the matrix is denoted with $\bold{X}_\star$. The inverse and transpose are written as $\mathbf{X}^{-1}$ and $\mathbf{X}^T$ respectively. The notation $\text{BDiag}(\bold{X}_i,i=1..N)$ forms a block-diagonal matrix with $\bold{X}_1$ to $\bold{X}_N$ located on the block-diagonal. For a block-diagonal matrix $\bold{X}$, the $i$-th block is denoted with $\bold{X}\left[i\right]$. \textsc{Matlab}\xspace-like notations are used to denote columns and rows of matrices, e.g $\mathbf{X}(:,\mcode{i})$ refers to the $i$-th column of $\mathbf{X}$, $\mathbf{X}(\mcode{i},:)$ the $i$-th row.
The vectorization operator applied on $\mathbf{X}$ is written with ${\rm vec}(\mathbf{X}) = \begin{bmatrix} x_{1,1} & x_{2,1} & \hdots & x_{m,n} \end{bmatrix}^T$. 
The operation of reshaping a vector into a matrix is denoted with $\text{ivec}$, e.g $\text{ivec}(\text{vec}(\bold{X})) = \bold{X}$. The Kronecker product of two matrices $\bold{X}, \bold{Y}$ is represented by the symbol $\otimes$ such as $\bold{X} \otimes \bold{Y}$. 
The Frobenius norm for a matrix $\mathbf{X} \in \mathbb{R}^{m \times n}$ is denoted with $\| \mathbf{X} \|_F^2 = \sqrt{\sum_{i = 1}^m\sum_{j = 1}^n x_{i,j}^2}$. The maximum singular value of $\mathbf{X}$ is denoted with $\lambda_{max}(\bold{X})$.
\newline
The big-$\mathcal{O}$ notation is used for describing computational complexities and indicates the asymptotic growth rate of the computational cost for a given mathematical operation. E.g an operation costing $\mathcal{O}(n)$ floating-point operations (flops) finishes in at most $c \cdot n$ flops, for some constant $c$.  
\newline
Other section-specific notations are introduced in the respective section.

\section{Preliminaries}
\label{sec11}
The main computational rules related to the Kronecker product are described in the appendix of this dissertation. In this section, we review some of the most important properties related to the decomposition of matrices with a sum of Kronecker products. Such a decomposition relies on the existence of block-matrices of equal size and that allow for a re-organization of the entries into a low-rank reshuffled matrix. 
\begin{defn} \cite{Loan:2000} \label{qks-def1}
Let $m_1,n_1,m_2,n_2 \in \mathbb{R}$. Let $\bold{X} \in \mathbb{R}^{m_1 m_2  \times n_1 n_2}$ and $\bold{X}_{i,j} \in \mathbb{R}^{m_2 \times n_2}$ such that:
\[
 \bold{X} = \left[ \begin{matrix} \bold{X}_{1,1} & \cdots & \bold{X}_{1,n_1} \\ \vdots & \ddots
     & \vdots \\ \bold{X}_{m_1,1} & \cdots & \bold{X}_{m_1,n_1} \end{matrix} \right]
\]
then the re-shuffle operator ${\mathcal R}(\bold{X}) \in \mathbb{R}^{m_1 n_1 \times m_2 n_2} $ is defined as:
\begin{equation}
 {\mathcal R}(\bold{X}) = \left[ \begin{matrix} {\rm vec}\big( \bold{X}_{1,1} \big)^T
     \\ \vdots \\  {\rm vec}\big( \bold{X}_{m_1,1} \big)^T \\ {\rm vec}\big(
     \bold{X}_{1,2} \big)^T \\ \vdots \\ {\rm vec}\big( \bold{X}_{m_1,n_1} \big)^T \end{matrix} \right]
\end{equation}
\end{defn}
\noindent
There exists a permutation matrix $\bold{P}$ in the set $\mathbb{R}^{m_1 n_1 m_2 n_2 \times m_1 n_1 m_2 n_2}$ such that:
\begin{equation}
\label{permutation}
\text{vec}(\mathcal{R}(\bold{X})) = \bold{P}\text{vec}(\bold{X})
\end{equation}
\begin{lem} \cite{Loan:2000} 
\label{qks-lem1}
Let $\bold{X} = \bold{F} \otimes \bold{G}$, with $(\bold{F},\bold{G}) \in \mathbb{R}^{m_1 \times n_1} \times \mathbb{R}^{m_2 \times n_2}$. Then:
\begin{equation}
  {\mathcal R}(\bold{X}) = {\rm vec} (\bold{F})  {\rm vec} (\bold{G})^T 
\end{equation} 
\end{lem}
\noindent
The operation in Lemma~\ref{qks-lem1} can also be reversed by the definition of the inverse vec operator ${\rm ivec}(.)$.
\begin{lem} \cite{Loan:2000} \label{qks-lem2}
Let $\bold{X}$ be defined as in Definition~\ref{qks-def1} and let an SVD of ${\mathcal R}(\bold{X})$ be given as:
\begin{equation}
 {\mathcal R}(\bold{X}) = \sum_{\ell =1}^{r} \sigma_\ell
 \bold{u}_\ell \bold{v}_\ell^T 
\end{equation} 
and let ${\rm ivec}\big( \bold{u}_\ell \big) = \bold{U}_\ell$, ${\rm ivec}\big( \bold{v}_\ell \big) = \bold{V}_\ell$, then:
\begin{equation}
  \bold{X} = \sum_{\ell =1}^{r}  \sigma_\ell \bold{U}_\ell \otimes \bold{V}_\ell 
\end{equation}
\end{lem}
\noindent
The integer $r$ is called the {\em Kronecker rank} of $\bold{X}$ with respect to the chosen block partitioning of $\bold{X}$ as given in Definition~\ref{qks-def1}. When $r$ is much smaller than $N$, $\bold{X}$ is said to have low-Kronecker rank. From Lemma~\ref{qks-lem2}, looking for a low-Kronecker rank approximation of a matrix is equivalent to finding a low-rank approximation of the reshuffled matrix. The reshuffling operator $\mathcal{R}$ as defined in Definition~\ref{qks-def1} that yields a reshuffled matrix of minimal rank $r$ is not unique: reshuffling the block-matrices row-wise rather than column-wise would yield the same Kronecker rank for $\bold{X}$. It then corresponds to the transpose of $\mathcal{R}(\bold{X})$.
\begin{defn} ($\alpha$-decomposable matrices, \cite{Mass:14}) \newline
Let us consider a network of subsystems such that the latter belong to $\alpha$ different classes, themselves composed of $N_i$ subsystems. Let $\mathbfcal{P} \in \mathbb{R}^{N \times N}$ be a pattern matrix. Define $\beta_j = \sum_{i=1}^j N_i$ (with $\beta_0 = 0$) and $\bold{I}_{\left[a_1:a_2\right]}$ as an $N \times N$ diagonal matrix which contains $1$ in the diagonal entries of indices from $a_1$ to $a_2$ (included) and $0$ elsewhere, then an $\alpha$-decomposable matrix (for a given $\alpha$) is a matrix of the following kind:
\[
 \mathbfcal{M} = \sum_{i=1}^\alpha \bigl(\bold{I}_{[\beta_{i-1}+1:\beta_i]}
 \otimes \bold{L}^{(i)} + \bold{I}_{[\beta_{i-1}+1:\beta_i]}
 \mathbfcal{P} \otimes \bold{N}^{(i)} \bigr)
\]
The matrices $\bold{L}^{(i)}$ are the diagonal blocks of $\mathbfcal{M}$ that model the local dynamics, while the influence from the neighborhood is represented by the matrices $\bold{N}^{(i)}$, according to the structure of $\mathbfcal{P}$.
\end{defn}
\noindent
When a state-transition matrix of a state-space model belongs to the class of $\alpha$-decomposable matrices, the associated network has a known interconnection pattern whose adjacency matrix is $\mathbfcal{P}$ while $\alpha$ represents the number of non-identical subsystems in the network. The pattern matrix $\mathbfcal{P}$  is allowed to be time-varying. For $\alpha = 1$ (and $\beta_1 = N$), these matrices are simply called {\em decomposable} matrices. 

As a generalization of this class of structured matrices, we define next the class of sums-of-Kronecker product matrices. 

\begin{defn}
The class of sums-of-Kronecker product matrices contains matrices of the following kind:
\[
\mathbfcal{M} = \sum_{i=1}^r \bold{M}_a^{(i)} \otimes \bold{M}_b^{(i)}
\]
with $\bold{M}_a^{(i)} \in \mathbb{R}^{m_1 \times n_1}$ and $\bold{M}_b^{(i)}  \in \mathbb{R}^{m_2 \times n_2}$. This class is denoted with $\mathcal{K}_{2,r}$. The matrices $\bold{M}_a^{(i)}, \bold{M}_b^{(i)}$ are called factor matrices.
\end{defn} 
\noindent
With this class of sums-of-Kronecker matrices, it is not necessary to have knowledge of a pattern matrix $\mathbfcal{P}$ as with decomposable matrices. Therefore, the topology of the network need not to be known in advance. Moreover, the network may be composed of heterogeneous subsystems without any further specifications on the structure of the factor matrices. When describing large-scale networks, this structure is advantageous for its high compression capabilities. While $m_1m_2n_1n_2$ entries are necessary to describe $\mathcal{M}$ in the unstructured case, only $r(m_1n_1+m_2n_2)$ elements are required in the sums-of-Kronecker framework. 

The next lemma provides insight on the benefits to use the class of Kronecker matrices to speed up simple linear algebra operations.
\begin{lem} \label{qks:lem4}
Let $\bold{x} \in \mathbb{R}^{N^2}$. Then, the orders of magnitude of the computational complexity orders for matrix-vector multiplication, matrix-matrix multiplication and matrix inversion is as follows:
\begin{center}
  \begin{tabular}{ l | c | c }
     & $\bold{A},\bold{B} \in \mathbb{R}^{N^2 \times N^2}$ & $\bold{A},\bold{B} \in \mathcal{K}_{2,r}$ \\ \hline
    $\bold{Ax}$  & $\mathcal{O}(N^4)$ & $\mathcal{O}(rN^3)$ \\ 
    $\bold{AB}$  & $\mathcal{O}(N^6)$ & $\mathcal{O}(r^2N^3)$ \\ 
    $\bold{A}^{-1}$ (case $r = 1$) & $\mathcal{O}(N^6)$ & $\mathcal{O}(N^3)$ \\ 
  \end{tabular}
\end{center} 
The complexity obtained with the Kronecker parametrization considers the operations required for forming the factor matrices only.
\end{lem}
\noindent
\noindent
\textbf{Proof.}
The matrix vector multiplication $\bold{Ax} = \bigl(\sum_{i = 1}^r \bold{M}_{\ell,i} \otimes \bold{M}_{r,i}\bigr) \bold{x}$ is rewritten into $\sum_{i = 1}^r \bold{M}_{r,i} \text{ivec}(\bold{X}) \bold{M}_{\ell,i}^T$. The complexity in the matrix format is $2rN^3$ compared to $N^4$ without exploiting the sums-of-Kronecker structure. 
When computing the matrix-matrix multiplication, only the products between factor matrices are computed yielding a cost of $r^2N^3$.
The inverse for $\bold{A}$ is determined via $\bold{A}^{-1} = \bold{A}_{\ell,1}^{-1} \otimes \bold{A}_{r,1}^{-1}$. Computing $\bold{A}_{\ell,1}^{-1}$ and $\bold{A}_{r,1}^{-1}$ costs $\mathcal{O}(N^3)$. 
\qed
\newline 
\begin{rem}
Approximating the inverse of large-scale low-Kronecker rank matrices $\bold{A} \in \mathcal{K}_{2,r}$ when the Kronecker rank is larger than one is an on-going research topic which \cite{beylkin} and \cite{giraldi} have investigated. 
\end{rem}
\noindent
From Lemma~\ref{qks:lem4}, efficient linear algebra operations are possible when $r$ is much smaller than $N$ which is the class of Kronecker models we are interested in.

\section{Problem formulation}
\label{sec12}
Low-Kronecker rank matrices are now used to model the input-output relationship of 2D networked systems. 

\subsection{QUARKS models}
Let us consider a regular grid with $N \times N$ nodes, each of which is associated with a scalar sensor signal. We assume $N$ strictly larger than $1$. Although the framework that we present here extends straightforwardly to arrays with nodes having multiple outputs, we only dwell on this case in Section~\ref{sec15}. The sensor readings at the time instant $k$ are stored in the matrix $\bold{S}(k)$ as:
\begin{equation}
 \label{StoreS}
 \bold{S}(k) = \left[ \begin{matrix} s_{1,1}(k) & s_{1,2}(k) & \cdots &
     s_{1,N}(k) \\ s_{2,1}(k) & s_{2,2}(k) &   & s_{2,N}(k) \\ \vdots & \vdots
     & \ddots & \vdots \\ 
     s_{N,1}(k) & s_{N,2}(k) & \cdots & s_{N,N}(k) \end{matrix} \right]  \in \mathbb{R}^{N \times N}
\end{equation}
In this paper we will consider that the temporal dynamics of this array of sensors is governed by the following VAR model:
\begin{equation}
 \label{VARX}
{\rm vec}\big( \bold{S}(k) \big)  = \sum_{i = 1}^p \bold{A}_i {\rm vec}\big( \bold{S}(k-i) \big) + {\rm vec}\big( \bold{E}(k) \big) 
\end{equation}
where ${\rm vec}\big( \bold{E}(k) \big)$ zero-mean white noise with identity covariance matrix. Covariance estimation for low-Kronecker rank matrices has been addressed in \cite{Tsi:13} and is not the subject of further investigations in this paper.
The coefficient matrices $\bold{A}_i$ in the VAR model \eqref{VARX} are in general highly structured. We consider the case they belong to the set ${\mathcal K}_r$ and focus on the coefficient matrices $\bold{A}_i$. To address an identification problem we parametrize these coefficient matrices as:
\begin{equation}
 \label{ParM}
\bold{A}_i = \sum_{j=1}^{r_i} \bold{A}_i^{(j)}, \qquad 
\bold{A}_i^{(j)} = \bold{M}(\bold{b}_i^{(j)})^T \otimes \bold{M}(\bold{a}_i^{(j)}) 
\end{equation}
with the vectors  $\bold{a}_i^{(j)}$ and  $\bold{b}_i^{(j)}$ parametrizing the
matrices $\bold{M}(\bold{a}_i^{(j)})$ and $\bold{M}(\bold{b}_i^{(j)})$ in an affine manner.
With the notation ${\rm vec}\Big( \bold{S}_k \Big) = \bold{s}_k$, the VAR model \eqref{VARX} can be rewritten as,
\begin{eqnarray}
 \label{ARXpk}
\bold{s}_k  & = &  \sum_{i=1}^p \Big( \sum_{j=1}^{r_i}
\bold{M}(\bold{b}_i^{(j)})^T \otimes \bold{M}(\bold{a}_i^{(j)}) \Big) \bold{s}_{k-i} + \bold{e}_k 
\end{eqnarray}
Using the following Kronecker rule, for matrices $\bold{X},\bold{Y},\bold{Z}$ of compatible dimensions such that the product $\bold{XYZ}$ exists,
\[
\Big( \bold{Z}^T \otimes \bold{X} \Big) {\rm vec}\Big(
\bold{Y} \Big) = {\rm vec}\Big( \bold{X Y Z} \Big)
\]
we can write the VAR model \eqref{ARXpk} as, 
 \begin{equation}
 \label{ARXp}
\bold{S}_k  = \sum_{i=1}^p \Big( \sum_{j=1}^{r_i}
\bold{M}(\bold{a}_i^{(j)}) \bold{S}_{k-i} 
\bold{M}(\bold{b}_i^{(j)}) \Big)  + \bold{E}_k
\end{equation}
This can also be written explicitly as,
 \begin{eqnarray}
 \label{ARXpe}
\bold{S}_k & = & \sum_{i=1}^p \bold{M}_{\bold{a}_i} \Bigl(\bold{I}_{r_i} \otimes \bold{S}_{k-i}\Bigr) \bold{M}_{\bold{b}_i} + \bold{E}_k
\end{eqnarray}
where 
\begin{eqnarray*}
\bold{M}_{\bold{a}_i} &=& \left[ \begin{matrix} 
\bold{M}(\bold{a}_i^{(1)}) & \cdots & \bold{M}(\bold{a}_i^{(r_i)}) \end{matrix}
\right]
\\
\bold{M}_{\bold{b}_i} &=& \left[ \begin{matrix} 
\bold{M}(\bold{b}_i^{(1)}) \\ \vdots \\ \bold{M}(\bold{b}_i^{(r_i)}) \end{matrix} \right] 
\end{eqnarray*}
The VAR(X) models \eqref{ARXpk}, \eqref{ARXp} or \eqref{ARXpe} are called {\em Kronecker VARX network models} and abbreviated with \emph{QUARKS models}.

\subsection{The identification problem of QUARKS models}
\label{sec_ideP}
Given the model structure of the QUARKS models, the problem of identifying these models from measurement sequences $\{ \bold{S}(k) \}_{k=1}^{N_{\rm t}}$ is fourfold:
\begin{enumerate}
 \item The temporal order index $p$.
 \item The spatial order index $r_i$ for each coefficient matrix. 
 \item The parametrization of the matrices $\bold{M}(\bold{a}_i^{(j)})$ and
   $\bold{M}(\bold{b}_i^{(j)})$. An example of a parametrization of the matrices $\bold{M}(\bold{a}_i^{(j)})$ and
   $\bold{M}(\bold{b}_i^{(j)})$ is (block) Toeplitz.
 \item The estimation of the parameter vectors $\bold{a}_i^{(j)}$,
   $\bold{b}_i^{(j)}$ up to an ambiguity transformation. This requires the specification of a cost function. An example of such a cost function using the model \eqref{ARXp} is the following least squares cost function,
\end{enumerate}
\begin{equation}
 \label{QUARKScost}
\min_{\bold{a}_i^{(j)},\bold{b}_i^{(j)}}  \sum_{k=p+1}^{N_{\rm t}}  \| \bold{S}(k)  - \sum_{i=1}^p \big( \sum_{j=1}^{r_i}
\bold{M}(\bold{a}_i^{(j)}) \bold{S}(k-i) 
\bold{M}(\bold{b}_i^{(j)}) \big)\|_F^2
\end{equation}
for data batches with $N_t$ points. 

By the selection of the parameter $p$ and the particular choices of the parametrization in step 3 above, various special cases of restricting the coefficient matrices $\bold{A}_i$ in \eqref{VARX} to particular sets such as ${\mathcal K}_{2,r_i}$ can be considered. Further constraints to the least-squares cost function \eqref{QUARKScost} might be introduced to look for sparsity in the parametrization vectors $\bold{a}_i^{(j)}$ and $\bold{b}_i^{(j)}$. 

The non-uniqueness of the optimal solution for the cost function \eqref{QUARKScost} is highlighted next. 
One way to solve this estimation problem is via vectorization of the sensor signals $\bold{S}(k)$:
\begin{eqnarray}
 \label{QUARKScost_vec}
\min_{\bold{a}_i^{(j)},\bold{b}_i^{(j)}} &&  \sum_{k=p+1}^{N_{\rm t}}  \| \bold{s}(k) - \sum_{i=1}^p \bold{A}_i \bold{s}(k-i) \|_2^2 \nonumber \\ 
\text{s.t} &&  \bold{A}_i = \sum_{j=1}^{r_i} \bold{M}(\bold{b}_i^{(j)})^T \otimes \bold{M}(\bold{a}_i^{(j)})
\end{eqnarray}
From \eqref{permutation}, the reshuffling operator $\mathcal{R}(.)$ is bijective in $\mathbb{R}^{N^2 \times N^2}$, therefore the above minimization problem is equivalent to: 
\begin{eqnarray}
\min_{\bold{a}_i^{(j)},\bold{b}_i^{(j)}} &&  \sum_{k=p+1}^{N_{\rm t}}  \| \bold{s}(k) - \sum_{i=1}^p \bold{A}_i \bold{s}(k-i) \|_2^2 \nonumber \\
\label{constraint_ambiguity}
\text{s.t} &&  \mathcal{R}(\bold{A}_i) = \bold{U}_i\bold{V}_i^T 
\end{eqnarray}
where:
\begin{eqnarray*}
\bold{U}_i &=& \begin{bmatrix}
\text{vec}\bigl(\bold{M}(\bold{a}_i^{(1)})\bigr) & \hdots & \text{vec}\bigl(\bold{M}(\bold{a}_i^{(r)})\bigr)
\end{bmatrix}  \\
\bold{V}_i &=& \begin{bmatrix}
\text{vec}\bigl(\bold{M}(\bold{b}_i^{(1)})\bigr) & \hdots & \text{vec}\bigl(\bold{M}(\bold{b}_i^{(r)})\bigr)
\end{bmatrix}
\end{eqnarray*}
For a non-singular transformation $\bold{T}_i \in \mathbb{R}^{r \times r}$, the constraint \eqref{constraint_ambiguity} can be equivalently written as:
\begin{equation}
\mathcal{R}(\bold{A}_i) = \overset{\sim}{\bold{U}}_i \overset{\sim}{\bold{V}}_i^T
\end{equation}
where: $\overset{\sim}{\bold{U}}_i = \bold{U}_i \bold{T}_i$ and $\overset{\sim}{\bold{V}}_i^T = \bold{T}_i^{-1} \bold{V}_i^T$. The non-uniqueness of the factor matrices is not an issue for practical use of QUARKS models as it does not affect the prediction-error.
\newline
\begin{rem} Let $m \in \{1,..,N^2\}$. Blind source separation \eqref{bss} as described in \cite{bss}  reshapes either (or both) the mixing vectors $\bold{M}(\mcode{m},:)$ and sources $\bold{S}(\mcode{m},:)$ in \eqref{bss} to form low-rank matrices. 
Then, there exists \emph{different} left and right matrices for each mixing vector $\bold{M}(\mcode{m},:)$ such that $\mathcal{R}(\bold{M}(\mcode{m},:)) = \bold{u}_m \bold{v}_m^T$, or equivalently,
\[
\bold{M}(\mcode{m},:) = \sum_{j = 1}^r \bold{u}_m(:,\mcode{j})^T\otimes \bold{v}_m(:,\mcode{j})^T
\]
where $\bold{u}_m \in \mathbb{R}^{I \times r}, \bold{v}_m \in \mathbb{R}^{J \times r}$ for two scalars $I,J$. The parameters $I,J$ are user-defined contrary to the QUARKS modeling, where $I,J = N$. Hence, all mixing vectors are decoupled independently contrary to the description for the QUARKS model \eqref{ParM} which assumes that the reshuffling into a matrix of \emph{both} the rows and columns of the mixing matrix $\bold{M}$ is low-rank.
\newline
We illustrate in the case where $p = 1$ and $\bold{M} = \bold{A}_1$. If $\text{rank}(\mathcal{R}(\bold{M})) = r$, then $\text{rank}(\mathcal{R}(\bold{M}(\mcode{m},:))) = r$ and $\text{rank}(\mathcal{R}(\bold{M}(:,\mcode{m}))) = r$. Fixing $I,J$ to $N$ and considering $N^2$ sources, there are $2rN^3$ unknown coefficients to estimate while the modeling \eqref{ParM} represents the coefficient matrices with $2rN^2$ entries. The QUARKS modeling decreases the data storage requirements by an order of magnitude.
\end{rem}
\noindent
An important challenge in solving the parameter estimation problem \eqref{QUARKScost} is the {\em computational efficiency} for the case when the size $N$ of the array is assumed to be large.

\section{Regularization inducing spatial-temporal stability and sparsity}
\label{sec13}
The Kronecker rank is assumed equal for all $i$, i.e $r_i = r$, without constraining the insights in this section. 

\subsection{Stability of VAR models}
In \cite{Chiuso:2012}, the stability for VAR models is guaranteed by modeling the impulse response from one node to all the other ones in the network as a zero-mean Gaussian process and with an adequately chosen covariance matrix, which ensures that the parameters of the impulse response are decaying with increasing temporal index. We refer to \cite{Hastie:08} for a general introduction to kernel methods and to \cite{Chen:2012} and \cite{Pillonetto:14} for an application to system identification.
In the following paragraph we integrate these results as an additional regularization to the cost function \eqref{QUARKScost}. 
We introduce the positive-definite matrix $\bold{P}_t \in \mathbb{R}^{p \times p}$ following a Gaussian-kernel to fit stable impulses. For example, a Diagonal-Correlated kernel $\bold{P}_t$ is defined with:
\begin{equation}
p_{t,(i,j)} = \xi^{\frac{i+j}{2}} \eta^{|i-j|} 
\end{equation}
for $i,j = 1..p$, and where the optimal hyperparameters $-1 \leq \eta \leq 1, 0 \leq \xi < 1$ shall be determined either by grid search or within the framework of Bayesian optimization to tune both the decay rate and the smoothness of the impulse response. Let $\bold{W}_t$ be a square root of $\bold{P}_t^{-1}$. 
As there is no prior information nor physical meaning to distinguish between the different factor matrices, these are regularized independently with the cost:
\begin{equation}
\sum_{j = 1}^r \| \bold{Q}_t \begin{bmatrix} \bold{U}_1(:,\mcode{j})\bold{V}_1(:,\mcode{j})^T \\ \vdots \\ \bold{U}_p(:,\mcode{j}) \bold{V}_p(:,\mcode{j})^T \end{bmatrix} \|_F^2
\end{equation}
where $\bold{Q}_t =  \bold{W}_t \otimes \bold{I}_{N^2} $. 
In a more compact notation, we write:
\begin{eqnarray*}
\bold{f}(\bold{M}_a^{(j)},\bold{M}_b^{(j)}) &=& \begin{bmatrix} \bold{U}_1(:,\mcode{j})\bold{V}_1(:,\mcode{j})^T \\ \vdots \\ \bold{U}_p(:,\mcode{j}) \bold{V}_p(:,\mcode{j})^T \end{bmatrix} \\
r_t(\bold{M}_a,\bold{M}_b) &=& \sum_{j = 1}^r \| \bold{Q}_t \bold{f}(\bold{M}_a^{(j)},\bold{M}_b^{(j)}) \|_F^2
\end{eqnarray*}
Such a regularization $r_t(.)$ is bilinear in the unknowns $\bold{M_a}_i^{(j)}, \bold{M_b}_i^{(j)}$.

\subsection{Spatial sparsity}
Real graphs or the regular networks from discretized Partial Differential Equations are such that each node is connected to a very limited number of other nodes with respect to the network's size. In the latter case, the neighborhood is localized which gives rise to a multi-banded structure of the full coefficient matrices, equivalent to a banded structure of each factor matrix. In case of high coupling, as is observed e.g in the atmospheric turbulence modeling as discussed in Section~\ref{sec15}, we rather tune the decay of the parameters away from the main diagonal rather than minimizing the number of non-zero entries. Furthermore, it will become clear in the next section that we would like to avoid all non-differentiable functions in the cost function, hence the focus is laid on kernel methods rather than on minimizing the $\ell_1$-norm of the factor-matrices. In this framework, an exponentially decreasing sequence has been studied in \cite{Chiuso:2012} for sparse network identification. 
We introduce a diagonal matrix $\bold{K}_s$ such that:
\begin{equation}
\bold{K}_s = \begin{bmatrix} \bold{I}_{N}k_{1} & 0 & \hdots & 0 \\
					  		 0 & \bold{I}_{2(N-1)}k_2 & \ddots & \vdots \\
					  		 \vdots & \ddots & \ddots & 0 \\
					  		 0 & \hdots & 0 & \bold{I}_{2}k_{N} \\
	  	     \end{bmatrix} \in \mathbb{R}^{N^2 \times N^2}
\end{equation}
where the scalars $k_{i}$ are such that $0 < k_i < k_{i+1}$. For example, a valid choice of such scalars is $k_i = e^{\zeta i}$ with $\zeta > 0$. 
Let $i \in \{0,\hdots,N-1\}$. For a matrix $\bold{X} \in \mathbb{R}^{N \times N}$, we denote with $\text{diag}(\bold{X},i)$ the $i$-th diagonal above the main diagonal and with $\text{diag}(\bold{X},-i)$ the $i$-th diagonal below the main diagonal. These vectors are then concatenated into a vector $\bold{d}_i$ defined with:
\[
\forall i \in \{1,\hdots,N-1\}, \bold{d}_i = \begin{bmatrix} \text{diag}(\bold{X},i)^T & \text{diag}(\bold{X},-i)^T \end{bmatrix}^T
\]
We reshape the elements of a square matrix diagonal-wise, starting by the main diagonal, and denote this operation with the operator $\mathcal{D}$:
\[
\mathcal{D}(\bold{X}) = \begin{bmatrix} \text{diag}(\bold{X},0)^T & \bold{d}_1^T & \hdots & \bold{d}_{N-1}^T \end{bmatrix}^T
\in \mathbb{R}^{N^2}
\]
The prior for matrices $\bold{M}(\bold{a}_i^{(j)}),\bold{M}(\bold{b}_i^{(j)})$ with values decaying away from the main diagonal is then:
\begin{equation}
\sum_{i = 1}^p \sum_{j = 1}^r \| \bold{K}_s \mathcal{D}\bigl(\bold{M}(\bold{a}_i^{(j)})\bigr) \mathcal{D}\bigl(\bold{M}(\bold{b}_i^{(j)})\bigr)^T \bold{K}_s^T \|_F^2
\end{equation}
This spatial regularization is denoted with $r_s(\bold{M}_\bold{a},\bold{M}_\bold{b})$.

\subsection{Structured factor matrices}
The parametrization of the factor matrices based on prior knowledge of the network may help either to further reduce the computational complexity of the model identification step, or to cast the model into a structure useful for control. The first category include banded, symmetric, Toeplitz and circulant patterns. Exploring such structures on the factor matrices is very attractive numerically as the number of parameters to be estimated reduces further. 

The block-Toeplitz Toeplitz-blocks structure arises e.g when modeling 2D homogeneous spatially-invariant phenomena on a rectangular grid. Many functions in optics are isotropic, for example the Point Spread Function or covariance matrix of the atmospheric turbulence, and can be modeled with a sum of few Kronecker terms. The Kronecker and block-Toeplitz Toeplitz-blocks structures are related, but not equivalent. 

\begin{lem} Let $\bold{X} \in \mathbb{R}^{N^2 \times N^2}$. 
\newline 
If $\bold{X}$ is symmetric block-Toeplitz, then $\bold{X}$ has a Kronecker rank at most equal to $N$. \newline 
If $\bold{X}$ has a Kronecker rank of one, it does not \emph{in general} imply neither that $\bold{X}$ is block-Toeplitz nor has Toeplitz-blocks.
\end{lem}
\noindent
\noindent
\textbf{Proof.} 
The first proposition is proved by using the reshuffling operator $\mathcal{R}$. It is then observed that the Toeplitz-blocks are not used in reducing further the Kronecker rank. 
\newline 
The factor matrices may be for example randomly generated. 
\qed \newline 

The second category contains for example the sparse (with unknown pattern of non-zero entries) or SSS structure. The SSS structure is more general than the Toeplitz, especially when it comes to model spatially-varying systems. The efficient use of SSS matrices has been thoroughly studied in \cite{riceTAC} while the extension to Multi-Level structures is an on-going research question. Modeling each factor matrix of the model as SSS enables significant improvements in the computational cost for future simple linear algebra operations. For example, the cost for standard matrix computations scales linearly with respect to the matrix size. For example, inverting a matrix $\bold{M}$ belonging to $\mathbb{R}^{N^2 \times N^2}$ written as $\bold{M} = \bold{M}_1 \otimes \bold{M}_2$ in which both $\bold{M}_1$, $\bold{M}_2$ have a SSS structure requires $\mathcal{O}(N)$ operations instead of $\mathcal{O}(N^6)$. Because such a parametrization for the matrices $\bold{M}(\bold{a}_i^{(j)}),\bold{M}(\bold{b}_i^{(j)})$ is not affine in the parameters $\bold{a}_i^{(j)},\bold{b}_i^{(j)}$, the identification of the SSS matrices is performed offline, i.e after having obtained an estimate for the non-parametrized $\bold{M}(\bold{a}_i^{(j)}),\bold{M}(\bold{b}_i^{(j)})$.

\subsection{The regularized cost function for QUARKS identification}
The cost function for the identification of sparse stable QUARKS models reads: 
\begin{eqnarray}
\label{CF}
\min_{\bold{a}_i^{(j)},\bold{b}_i^{(j)}} &&   \sum_{k=p+1}^{N_{\rm t}} \| \bold{S}(k)  - \sum_{i=1}^p \big( \sum_{j=1}^{r_i}
\bold{M}(\bold{a}_i^{(j)}) \bold{S}(k-i) 
\bold{M}(\bold{b}_i^{(j)}) \big) \|_F^2 \nonumber\\
&& +\mu \cdot r_t(\bold{M}_a,\bold{M}_b) +\lambda \cdot r_s(\bold{M}_a,\bold{M}_b) 
\end{eqnarray}
where $\mu,\lambda$ are regularization parameters. The cost function \eqref{CF} belongs to the class of multi-convex problems in which fixing one set of variables yields a convex problem. Adding regularization to the cost function aims at decreasing the prediction error of the estimated VAR model when dealing with noisy and short data batches rather than speeding up the convergence as done in \cite{LiRALS}.

\begin{rem} The regularization in \eqref{CF} is \emph{bilinear} contrary to the one analyzed in \cite{Boyd:15}, \cite{Hornik}  within the framework of Principal Component Analysis (PCA). Based on \cite{Boyd:15}, a regularization for the minimization \eqref{constraint_ambiguity}  would minimize a (weighted) sum of the Frobenius norm of the factor matrices. 
\end{rem}

\section{Bi-convex Cost function approach}
\label{sec14}
The factor matrices are assumed unstructured in the upcoming sections. 

\subsection{An Alternating Least Squares approach}

The regularized least-squares representation \eqref{CF} is bilinear in its unknowns but features relatively small factor matrices, which has the advantage that constraints on the parametrization of the matrices $\bold{M}(\bold{a}_i^{(j)})$ and $\bold{M}(\bold{b}_i^{(j)})$ can be more easily taken into consideration than via a low-rank minimization on the large-scale reshuffled matrix. 
A non-linear optimization scheme such as the separable least-squares in \cite{bruls} proceeds with two steps, one of which however consists of non-linear optimization. Iterative algorithms have been derived as a generalization of the linear Gauss-Seidel iterations for solving coupled Sylvester matrix equations in \cite{ding}. Similarly as in \cite{hoff}, we propose to address \eqref{QUARKScost_vec} by solving a sequence of linear least-squares and using ALS, which is a special case of the block \emph{non-linear} Gauss-Seidel method as highlighted in \cite{LiRALS}. 
\newline
The data-fitting term in \eqref{CF} is first rewritten with:
\begin{equation}
\| \bold{\overset{\sim}{S}} - \bold{\overline{M}}_a \bold{X_b} \|_F^2
\end{equation}
where:
\[
\bold{\overset{\sim}{S}} = \begin{bmatrix} \bold{\overset{\sim}{S}}_{1,1} & \hdots & \bold{\overset{\sim}{S}}_{1,N} \\ \vdots & & 
\vdots \\ \bold{\overset{\sim}{S}}_{N,1} & \hdots & \bold{\overset{\sim}{S}}_{N,N} \end{bmatrix}, \quad
\bold{\overset{\sim}{S}}_{j,i} =  \begin{bmatrix} s_{j,i}(p+1) \\ \vdots \\ s_{j,i}(N_t) \end{bmatrix}
\]
We denote the $\ell$-th column of $\bold{M}(\bold{a}_i^{(j)})$ and $\bold{M}(\bold{b}_i^{(j)})$ with respectively $\bold{a}_{i,\ell}^{(j)}$ and $\bold{b}_{i,\ell}^{(j)}$.
\begin{eqnarray*}
\bold{\overline{M}}_a &=& \begin{bmatrix} \bold{\overline{M}}_{a,1} & \hdots & \bold{\overline{M}}_{a,p} \end{bmatrix}  \\ 
\bold{\overline{M}}_{a,i} &=& \begin{bmatrix} \bold{\overline{M}}_{a,i,1} & \hdots & \bold{\overline{M}}_{a,i,r} \end{bmatrix} \\
\bold{\overline{M}}_{a,i,j} &=& 
(\bold{I}_N \otimes \bold{\overset{\sim} U}_i)
\begin{bmatrix} \bold{a}_{i,1}^{(j)} \otimes \bold{I}_N \\ \vdots \\ \bold{a}_{i,N}^{(j)} \otimes \bold{I}_N \end{bmatrix} \\
\bold{\overset{\sim} U}_i &=&  \begin{bmatrix} \bold{S}(p+1-i)(\mcode{1},:) & \hdots & \bold{S}(p+1-i)(\mcode{N},:) \\ \vdots & & \vdots \\ \bold{S}(N_t-i)(\mcode{1},:) & \hdots & \bold{S}(N_t-i)(\mcode{N},:) \end{bmatrix} \\
\bold{X_b} &=& \begin{bmatrix} \bold{M}_{\bold{b}_1}^T & \hdots & \bold{M}_{\bold{b}_p}^T \end{bmatrix}^T%, \quad \bold{x_b} = \text{vec}(\bold{X_b})
\end{eqnarray*} 
The term $\mu \cdot r_t(\bold{M}_a,\bold{M}_b)$ is rewritten as $\|\bold{F}_b(\bold{M}_a)  \bold{X_b} \|_F^2$, where $\bold{F}_b(\bold{M}_a)$ is a $p \times p$ block-matrix. The block at position $(i,j)$ is equal to:
\[
\sqrt{\mu} w_{t,(i,j)} \text{BDiag}(\bold{I}_N \otimes \text{vec}(\bold{M}(\bold{a}_i^{(m)})),m=1..r) 
\]
Moreover, a matrix $\bold{G}_b(\bold{M}_a)$ is derived such that the regularization for spatial sparsity reads:
\begin{equation}
\lambda \cdot r_s(\bold{M}_a,\bold{M}_b) = \| \bold{G}_b(\bold{M}_a)\text{vec}(\bold{X}_b)  \|_2^2
\end{equation}
where:
\begin{eqnarray*}
\bold{G}_b(\bold{M}_a) &=& \sqrt{\lambda}\bold{P}_{r,s}\text{BDiag}(\bold{G}_{b,j}(\bold{M}_a),j=1..r)\bold{P}_{c,s} \\
\bold{G}_{b,j}(\bold{M}_a) &=& \text{BDiag}(\bold{K}_s \otimes \bold{K}_s \mathcal{D}(\bold{M(a}_i^{(j)}) ,i = 1..p)
\end{eqnarray*}
The matrices $\bold{P}_{r,s}$ and $\bold{P}_{c,s}$ permute respectively the rows and columns such that $\bold{G}_b(\bold{M}_a)$ is block-diagonal. We denote the $i$-block in the main block-diagonal with $\bold{G}_b(\bold{M}_a) \left[i\right]$. The cost function \eqref{CF} is then separable for each column of $\bold{X_b}$:
\begin{equation}
\label{updateA}
\min_{\bold{X_b}} \quad \sum_{i = 1}^N \| \underbrace{\begin{bmatrix} \bold{\overset{\sim}{S}} \\ \bold{0} \\ \bold{0} \end{bmatrix}}_{\bold{Y}} - \underbrace{\begin{bmatrix} \bold{\overline{M}}_a \\ \bold{F}_b(\bold{M}_a) \\ \bold{G}_b(\bold{M}_a)\left[i\right] \end{bmatrix}}_{\bold{F_b}_i} \bold{X_b}(\mcode{:,i}) \|_F^2 
\end{equation}
Similarly, the least-squares for updating $\bold{X_a} = \begin{bmatrix} \bold{M}_{\bold{a}_1}^T & \hdots & \bold{M}_{\bold{a}_p}^T \end{bmatrix}^T$ is:
\begin{equation}
\label{updateB}
\min_{\bold{X_a}} \quad \sum_{i = 1}^N \| \bold{Y} - \underbrace{\begin{bmatrix} \bold{\overline{M}}_b \\ \bold{F}_a(\bold{M}_b) \\ \bold{G}_a(\bold{M}_b)\left[i\right] \end{bmatrix}}_{\bold{F_a}_i} \bold{X_a}(\mcode{:,i}) \|_F^2 
\end{equation}
where:
\begin{eqnarray*}
\label{mb}
\bold{\overline{M}}_b &=& \begin{bmatrix} \bold{\overline{M}}_{b,1,1} & \hdots & \bold{\overline{M}}_{b,1,r} & \hdots & \bold{\overline{M}}_{b,p,r}  \end{bmatrix} \\
\bold{\overline{M}}_{b,i,j} &=& 
(\bold{I}_N \otimes \bold{\overset{\sim} U}_i)
\begin{bmatrix} \bold{I}_N \otimes \bold{b}_{i,1}^{(j)} \\ \vdots \\ \bold{I}_N \otimes \bold{b}_{i,N}^{(j)}  \end{bmatrix} 
\end{eqnarray*}
%The solution is denoted with $\bold{X_a}^{(\kappa)}$.
The least-squares \eqref{updateA} and \eqref{updateB} are iteratively solved starting with some random initial guess for $\bold{X_a}$ until some stopping criterion is reached. 
The iterations are stopped when the decrease between two consecutive values of the cost function is lower than a given threshold. Algorithm~\ref{qks-alg} summarizes the steps. 

\begin{algorithm}
\DontPrintSemicolon
\SetAlgoLined
\SetKwInOut{Input}{Input}\SetKwInOut{Output}{Output}
\Input{$\{\bold{S}(k)\},r,p,\text{operators}(\bold{G}_a,\bold{F}_a,\bold{G}_b,\bold{F}_b) ,\{\| \bold{X_b}(:,\mcode{i}) \|_2 \}_{1..N}$  }
\Output{ $ \{ \widehat{\bold{M}}_{a_i}, \widehat{\bold{M}}_{b_i} \}_{i=1..p}$  }
\BlankLine 
\tcc{Default values} 
$ \kappa = 1, \kappa_{max} = 50, \epsilon = \infty, \epsilon_{min} = 10^{-3}$ \;
\tcc{Initial guesses}
$\bold{X_a}^{(0)} = \mcode{randn(Nrp,N)}$ \;
Form $\bold{\overset{\sim}{S}}$ and $\bold{Y}$ \;
\tcc{Start ALS}
\While{$\kappa < \kappa_{max}$ \text{and} $\epsilon > \epsilon_{min}$} {
\tcc{Optimize over $\bold{X_b}$}
Compute $\bold{F}_0^T\bold{F}_0$ where: $\bold{F}_0 := \begin{bmatrix} \bold{\overline{M}}_a^{(\kappa-1)} \\ \bold{F}_b(\bold{M}_a^{(\kappa-1)}) \end{bmatrix}$ \;
\For{$i = 1..N$}{
Form $\bold{F_b}_i$ \;
$\bold{X_b}^{(\kappa)}(:,\mcode{i}) := (\bold{F_b}_i^T\bold{F_b}_i)^{-1}\bold{F_b}_i^T \bold{Y}(:,\mcode{i})$ \;
\tcc{Normalize (if $r = 1$ and the true values of the norm of each column is available)}
\If {$\mcode{isempty}(\{\| \bold{X_b}(:,\mcode{i}) \|_2 \}_{1..N}) = 0$}{
$\bold{X_{b,n}}^{(\kappa)}(:,\mcode{i}) = \bold{X_{b}}^{(\kappa)}(:,\mcode{i})\frac{\| \bold{X_b}(:,\mcode{i}) \|_2}{\| \bold{X_{b}}^{(\kappa)}(:,\mcode{i}) \|_2}$ \;
}
}
\tcc{Optimize over $\bold{X_a}$}
Compute $\bold{F}_0^T \bold{F}_0$ where $\bold{F}_0 := \begin{bmatrix} \bold{\overline{M}_{b,n}}^{(\kappa)} \\ \bold{F}_a(\bold{M_{b,n}}^{(\kappa)}) \end{bmatrix}$ \;
\For {$i = 1..N$}{
Form $\bold{F_a}_i$ \;
$\bold{X_a}^{(\kappa)}(:,\mcode{i}) := (\bold{F_a}_i^T\bold{F_a}_i)^{-1}\bold{F_a}_i^T \bold{Y}(:,\mcode{i})$ \;
}
\tcc{Check stopping criterion}
$c^{(\kappa)} := \| \bold{F}\bold{X_a}^{(\kappa)} - \bold{Y} \|_F^2$ \;
$\epsilon = |c^{(\kappa)}-c^{(\kappa-1)} |$ \;
 $\kappa = \kappa+1$ \;
 }
\tcc{(useful for retrieving the only solution when the true values of the norm of each column is available)}
\For {$i = 1..p$}{
$\widehat{\bold{M}}_{b_i} = \bold{M}_{b_i}^{(\kappa-1)}\text{sign}(m_{b,i,(1,1)}^{(\kappa-1)}), 
\widehat{\bold{M}}_{a_i} = \bold{M}_{a_i}^{(\kappa-1)}\text{sign}(m_{b,i,(1,1)}^{(\kappa-1)})$ \;
}
\caption{ALS for QUARKS identification} \label{qks-alg}
\end{algorithm}

\begin{rem}
Normalization of the columns $\bold{X_{b}}^{(\kappa)}(:,\mcode{i})$ for $i$ in the set $\{1,..,N\}$ in line 9 of Algorithm~\ref{qks-alg} and the scaling in line 22 are added. It is considered within the scope of this paper for two reasons. First, this normalization is a key ingredient in deriving that the iterates converge to a fixed point. It however requires the knowledge of $\bold{X_{b}}(:,\mcode{i})$ which is rarely available in practice. Second, the non-uniqueness of the solution may imply that the estimated factor matrices scale to very large (respectively very low) values which a normalization prevents from happening. Numerical examples in Section~\ref{sec15} illustrate its impact on the convergence.
\end{rem}

\begin{rem} \emph{Initialization}. We highlight the importance of choosing random initial guesses to converge to a global minimum. For illustration, we choose $p = 1$ and $r > 1$. Then, for $\alpha \in \mathbb{N}, \alpha < r$, if the rank of $\bold{X}_a^{(0)}$ is $\alpha N$, we have observed that the solution to Algorithm~\ref{qks-alg} without normalization corresponds to the same solution that would have been obtained by choosing rather $r = \alpha$.
\end{rem}

\subsection{Convergence proof for the normalized ALS}
The musings in \cite{musings} detail properties about ALS and multilinear fittings in general. The solution to the QUARKS without normalization is not unique because of the ambiguity transformation and there are therefore infinitely many stationary points. The convergence of the global matrices $\{\bold{A}_i\}_{i=1..p}$ is a necessary condition but not sufficient for stopping the iterations. Again because of the non-uniqueness, the factor matrices might still change and compensate each other without modifying the value of the cost function. Whether the entries of the factor matrices converge to some value is a more adequate question. 

In this paragraph, we consider the normalized version of Algorithm~\ref{qks-alg} and prove that the iterations converge to a fixed point of a particular functional. We assume the temporal order $p$ and spatial order $r$ to be both equal to one and both regularization parameters equal to zero. 
The convergence proof relies on the work in \cite{Li:15} where the result is established when the unknowns are vectors. We review the results in the following for completeness and highlight the non-trivial extensions in the appendix of this paper. The convergence proof uses the Contraction Mapping Theorem, \cite{bookFP}. 

With $p,r$ equal to one, the columns of $\bold{X_a}$ are $\bold{a}_{1,i}^{(1)}$.
We abbreviate with $\bold{a}$ the vector concatenating all columns $\bold{a}_{1,i}^{(1)}$; the latter is abbreviated with $\bold{a}_i$. The estimate of $\bold{a}$ at iteration $\kappa$ is denoted with $\widehat{\bold{a}}^{(\kappa)}$. Similar notations hold for $\bold{b}$. A functional representation of the three steps in Algorithm~\ref{qks-alg} reads:
\begin{eqnarray}
\label{qks-iter1}
\widehat{\bold{b}}^{(\kappa)} &=& \mathcal{F}_1(\widehat{\bold{a}}^{(\kappa-1)}) \\
\widehat{\bold{b_n}}^{(\kappa)} &=& \mathcal{F}_2(\widehat{\bold{b}}^{(\kappa)}) \\
\label{qks-iter3}
\widehat{\bold{a}}^{(\kappa)} &=& \mathcal{F}_3(\widehat{\bold{b_n}}^{(\kappa)})
\end{eqnarray}
These equations can be expressed using a single operator $\mathcal{F}(.)$ mapping the estimate $\widehat{\bold{a}}^{(\kappa-1)}$ to $\widehat{\bold{a}}^{(\kappa)}$:
\begin{equation}
\widehat{\bold{a}}^{(\kappa)} = \mathcal{F}_3(\mathcal{F}_2(\mathcal{F}_1(\widehat{\bold{a}}^{(\kappa-1)}))) 
= \mathcal{F}(\widehat{\bold{a}}^{(\kappa-1)}) 
\end{equation}
\begin{lem} \emph{[The Contraction Mapping Theorem, \cite{bookFP}]} Let $(X, D)$ be a non-empty complete metric space where $D$ is a metric on $X$. Let $\mathcal{F} : X \rightarrow X$ be a contraction mapping on $X$, i.e., there is a non-negative real number $Q < 1$ such that $D(\mathcal{F} (\bold{x}), \mathcal{F} (\bold{y})) \leq Q D(\bold{x}, \bold{y})$, for all $\bold{x}, \bold{y} \in X$. Then the map $\mathcal{F}$ admits one and only one fixed point $\bold{x}^\star \in X$ which means $\bold{x}^\star - \mathcal{F}(\bold{x}^\star) = 0$. 
Furthermore, this fixed point can be found from the convergence of an iterative sequence defined by $\bold{x}^{(\kappa + 1)} = \mathcal{F} \bigl(\bold{x}^{(\kappa)}\bigr)$ for $k = 1, 2, ...$ with an arbitrary starting point $\bold{x}^{(0)}$ in $X$.
\end{lem}
\noindent
A fixed point of $\mathcal{F}$ is a stationary point of the cost function in the minimization \eqref{CF}. The reverse implication is not necessarily true: there are many other stationary points that are discarded from the analysis when normalizing. There are an infinite number of solutions all equivalent as all globally minimizing the cost function. Moreover, if the fixed point is unique as we show in this very particular case of ALS, it corresponds to the targeted factor matrices for which the norm of the columns is assumed to be known. We refer to these as the true values.

We now define a set associated to the true value $\bold{a}$:
\begin{eqnarray*}
%X_{\Br} &=& \{ \widehat{\Br} \in \mathbb{R}^{N^2} |  \forall i \in \{1,...,N\}, \| \widehat{\bold{b}_i} \|_2 = \| \bold{b}_i \|_2,  \widehat{b}_1 > 0   \} \\
X_{\Bell} &=& \{ \widehat \Bell \in \mathbb{R}^{N^2}  | \forall i \in \{1,...,N\}, \| \widehat{\bold{a}_i} \|_2 \leq \| \bold{a}_i \|_2  \}
\end{eqnarray*}
\begin{thm} \label{thm-qks}
Let $p = 1, r = 1$ and $(\lambda, \mu) = (0,0)$. \newline
If the following statements are true:
\begin{itemize}
\item $\bold{A1}$: the noise components in $\text{vec}(\bold{E}(k))$ are independent identically distributed (i.i.d) with zero-mean and finite variance. 
\item $\bold{A2}$: the matrix $\bold{\overset{\sim} U}_1$ is full column rank.
\item $\bold{A3}$: either $\| \bold{b}_i \|_2$ or $\| \bold{a}_i \|_2$ is known for all $i$ and the first non-zero entry of $\bold{b}$ or $\bold{a}$ is strictly positive.
\item $\bold{A4}$: the initial guess $\widehat{\bold{a}}^{(0)}$ is non-zero. 
\end{itemize}
Then, the map $\mathcal{F}: X_{\Bell} \rightarrow X_{\Bell}$ is a contraction on $X_{\Bell}$ when $N_t \rightarrow \infty$ and has a unique fixed point which corresponds to the true parameters $\Bell$.
\end{thm}
\noindent
\textbf{Proof.}
The proof is derived in the appendix of this paper.
\qed
\newline
The assumption \textbf{A2} corresponds to the persistency of excitation from the data and is a key ingredient in the convergence. When using ALS for a system identification problem and assuming $\bold{A2}$ enables to avoid rank-deficiencies in the matrix $\bold{F}$ and therefore swamps as observed for tensor decomposition in \cite{LiRALS} do not occur. 

Theorem~\ref{thm-qks} proves that whatever the non-zero initial conditions the iterations \eqref{qks-iter1} to \eqref{qks-iter3} converge to a fixed point asymptotically when $N_t$ approaches infinity. When the temporal order is strictly larger than one, the solution to an update in line 8 or 16 in Algorithm~\ref{qks-alg} is unique if and only if the matrix $\bold{F}$ is full column rank. This condition provides with indications on how to choose the initial guess. In practice, we choose randomly generated initial guesses independent for each factor matrix such that $\bold{X_a}^{(0)}$ is full column rank. 

%\begin{rem} \emph{Larger Kronecker ranks}. For both $(p,r)$ strictly greater than one, and if the assumptions $\bold{A1}$ and $\bold{A2}$ are satisfied and if the initial guess $\bold{M}_a^{(0)}$ is full rank, global convergence was observed for all numerical experiments carried out with the non-normalized version of Algorithm~\ref{qks-alg}, i.e skipping lines 9 to 11. If the Kronecker rank is larger than one, we have observed numerically that the normalized version of Algorithm~\ref{qks-alg} requires a lot more iterations to converge to a global minimum than without normalization. 
%\end{rem}

\subsection{Computational complexity}
\subsubsection{Unstructured VAR}
For matters of comparison, we evaluate first the complexity for estimating the coefficient matrices associated with the unstructured VAR. Using \eqref{VARX} with temporal data within the range $\{1,...,N_t\}$ with $N_t \geq N^2p$ to recover a unique solution, we write:
\begin{eqnarray*}
\bold{S_f} &=& \begin{bmatrix} \bold{A}_1 & \hdots & \bold{A}_p \end{bmatrix} \bold{S_p} +  \bold{E_p} 
\end{eqnarray*}
where:
\begin{eqnarray*}
\bold{S_f} &=& \begin{bmatrix} \bold{s}(p+1) & \hdots & \bold{s}(N_t) \end{bmatrix} \\
\bold{S_p} &=& \begin{bmatrix} \bold{s}(p) & \hdots & \bold{s}(N_t-1) \\ \vdots & & \vdots \\ \bold{s}(1) & \hdots & \bold{s}(N_t-p) \end{bmatrix}
\end{eqnarray*}
The least-squares estimation for the coefficient matrices is hence equal to:
\begin{equation}
\begin{bmatrix} \widehat{\bold{A}}_1 & \hdots & \widehat{\bold{A}}_p \end{bmatrix}
=
\bold{S_f}\bold{S_p}^T\bigl( \bold{S_p} \bold{S_p}^T \bigr)^{-1}
\end{equation}
The complexity is summarized in Table~\ref{qks:tbl1}. The dependency on the number of temporal samples is kept: a correct identification in noisy conditions often requires $N_t \geq N^2p$. The complexity for estimating unstructured VAR is $\mathcal{O}(N^4N_t)$. 

\begin{table}[h!]
	\centering
    \begin{tabularx}{0.2\textwidth}{c  c}
    \hline   
	\textbf{Operation} & \textbf{Flops} \\ 
	\hline
    $\bold{ S_p} \bold{ S_p}^T $ & $\mathcal{O}(N^4N_t)$ \\  
    $(\bold{ S_p} \bold{ S_p}^T)^{-1}$ & $\mathcal{O}(N^6)$ \\ 
    $\bold{ S_f} \bold{ S_p}^T $ & $\mathcal{O}(N^4N_t)$ \\  
    \hline
    \end{tabularx}
        \caption{Computational complexity for the unstructured estimation of a VAR model.}
        \label{qks:tbl1}
\end{table}

\subsubsection{QUARKS}
We assume that the Kronecker rank and the number of iterations to reach convergence are independent of $N$. In practice, larger arrays require a larger number of temporal samples and therefore $N_t$ is included in the computational count. The lines 5, 8, 13, 16 are the most computationally costly of Algorithm~\ref{qks-alg}. There are two case worth investigating: $\lambda = 0$ and $\lambda \neq 0$. 

If $\lambda = 0$, the pseudo-inverse for the matrix $\bold{F}_0$ is computed only once at each iteration. Forming $\bold{\overline{M}_a}^{(\kappa-1)}$ requires $(N_t-p)rp$ matrix-matrix multiplications of size $N \times N$. The number of temporal samples is such that $N(N_t-p) \geq Nrp$ to guarantee a unique solution of each subproblem without regularization. The complexity is $\mathcal{O}(N^3N_t)$ flops.
Computing its inverse requires $\mathcal{O}(N^3)$ whereas right-multiplying the latter with $\bold{F}^T$ reaches $\mathcal{O}(N^3N_t)$. The computational complexity for Algorithm~\ref{qks-alg} with $\lambda = 0$ reaches $\mathcal{O}(N^3N_t)$ where $N_t \gg rp$. 

If $\lambda \neq 0$, the matrix $\bold{F}$ is partitioned in 3 parts. The cost for computing $\bold{F^TF}$ in line 9 boils down to computing $\bold{F}_0^T\bold{F}_0$ because $\bold{G}_b(\bold{M}_a^{(\kappa-1)})\left[i\right] $ is sparse. Moreover, the term $\bold{F}_0^T\bold{F}_0$ is computed once in line 5 with $\mathcal{O}(N^3N_t)$ flops.
Computing the inverse of $\bold{F}^T\bold{F}$ requires $\mathcal{O}(N^3)$ flops, while multiplying the inverted matrix with $\bold{F}^T$ costs $\mathcal{O}(N^3N_t)$. These two operations need to be repeated $N$ times, although it should be performed in parallel.
The price for computing the lines 13 and 16 is similar to the above discussion. When the algorithm is computed sequentially and without making use of distributed computing platforms, the overall cost reaches $\mathcal{O}(N^4N_t)$.

The costs are summarized in Table~\ref{qks:tbl2}. 

\begin{table}[h!]
	\centering
    \begin{tabularx}{0.3\textwidth}{l c}
    \hline     
	\textbf{Operation} & \textbf{Flops} \\ \hline 
    Lines 5 and 13 & $\mathcal{O}(N^3N_t)$ \\  
    Lines 8 and 16 (for each $i$) & $\mathcal{O}(N^3N_t)$ \\ 
    Lines 8 and 16 (total for all $i$) & $\mathcal{O}(N^4N_t)$ \\  
    Total (with $\lambda \neq 0)$ & $\mathcal{O}(N^4N_t)$ \\  
    Total (with $\lambda = 0)$ & $\mathcal{O}(N^3N_t)$ \\  
    \hline
    \end{tabularx}
        \caption{Computational complexity for the estimation of a QUARKS.}
        \label{qks:tbl2}
\end{table}

\section{Numerical examples}
\label{sec15}
The proposed QUARKS identification method is first illustrated with a randomly generated VARX model and then with an application to AO.

\subsection{Case study 1: Randomly generated VARX model}
We first illustrate the convergence of Algorithm~\ref{qks-alg} with different normalizations for a randomly generated QUARKS model whose temporal order and Kronecker rank is known. 
The model structure is the following:
\[
\bold{S}(k) = \sum_{i = 1}^{p} \sum_{j = 1}^r \bold{M}(\bold{a}_i^{(j)}) \bold{U}_{k-i} \bold{M}(\bold{b}_i^{(j)})
\]
where $N = 10, \bold{S}(k) \in \mathbb{R}^{10 \times 10}$. The factor matrices $\bold{M}(\bold{a}_i^{(j)})$ and $\bold{M}(\bold{b}_i^{(j)})$ are generated with a Toeplitz pattern and such that its entries decay away from the diagonal. The input is a white Gaussian noise. The number of temporal samples $N_t$ is set to $100 \times Npr$. Two scenarios were tested to analyze the influence of the temporal and spatial order $(p,r)$ on the convergence.

\begin{figure}[htbp]
    \centering
    \begin{minipage}{.49\textwidth}
        \centering
        \includegraphics[width=0.95\linewidth]{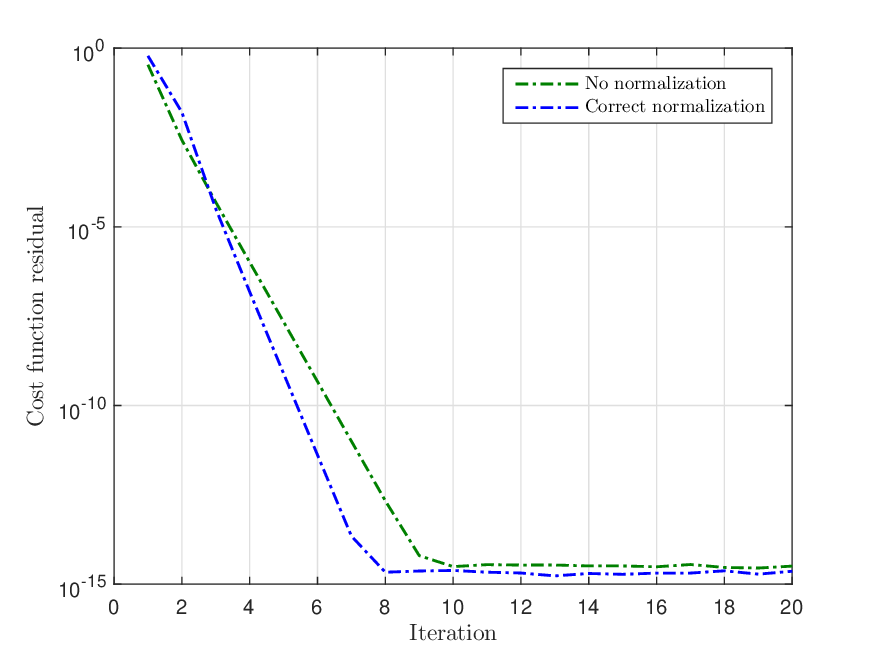}
        \\ (a)
    \end{minipage} 
    \begin{minipage}{0.49\textwidth}
        \centering
        \includegraphics[width=0.95\linewidth]{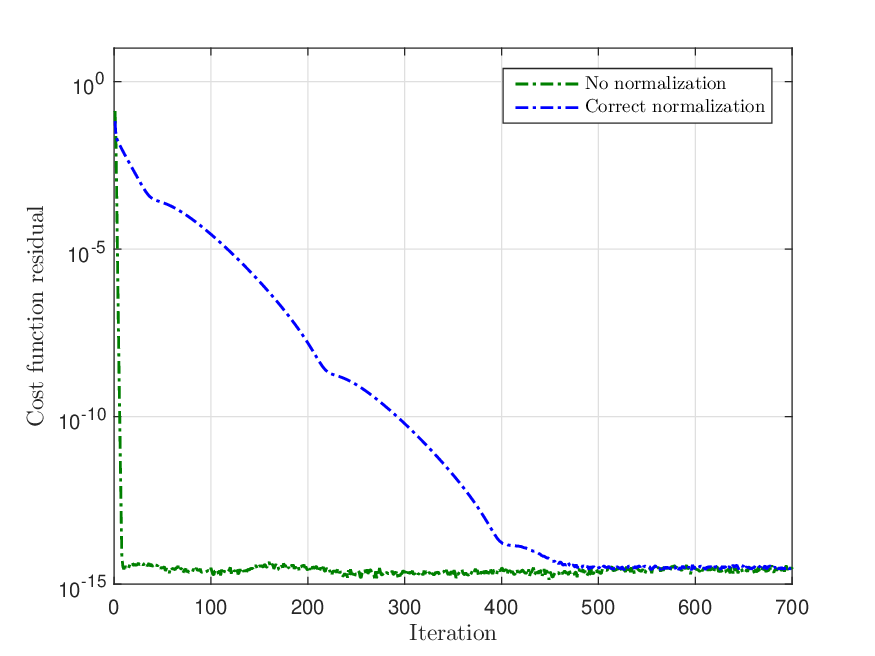}
        \\ (b)
    \end{minipage}
    \caption{Evolution of the cost function as a function of the number of iterations with two normalizations (no normalization, normalization as in Algorithm~\ref{qks-alg}). (a): the pair $(p,r)$ is set to $(2,1)$. (b): the pair $(p,r)$ is set to $(1,2)$.}
    \label{f1}
\end{figure}

\begin{figure}[htbp]
    \centering
    \begin{minipage}{.49\textwidth}
        \centering
        \includegraphics[width=0.95\linewidth]{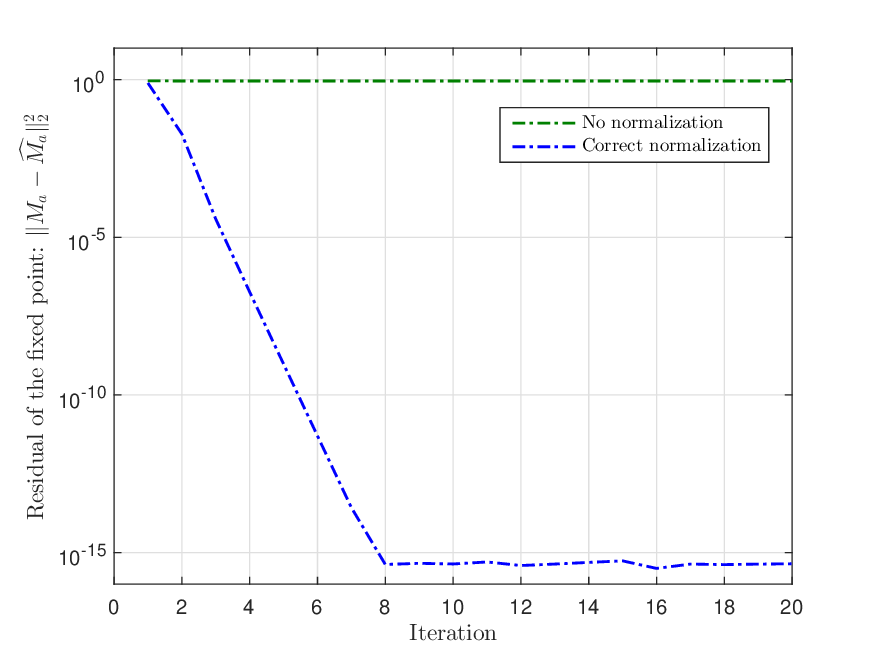} \\ (a)
    \end{minipage}
    \begin{minipage}{0.49\textwidth}
        \centering
        \includegraphics[width=0.95\linewidth]{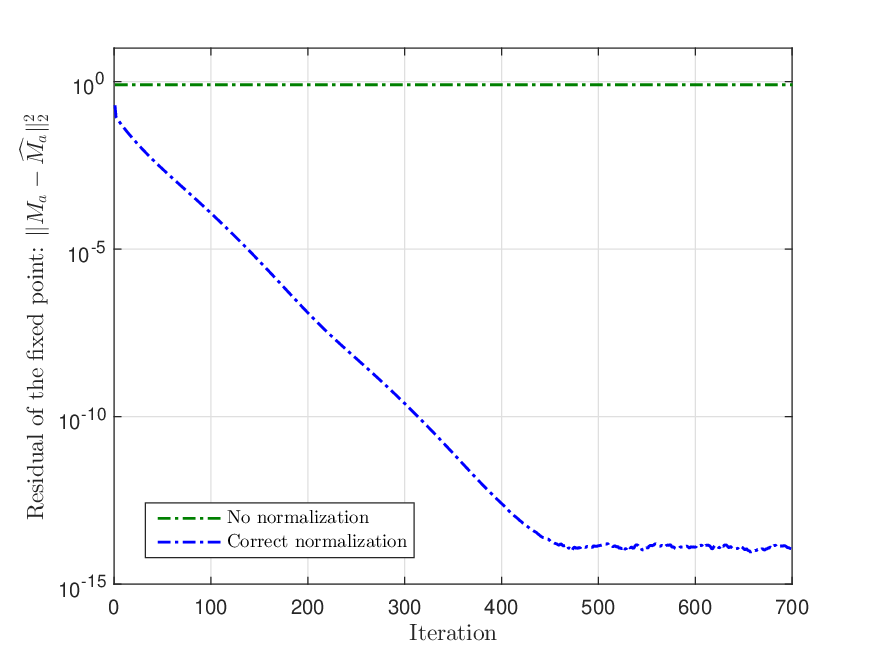} \\ (b)
    \end{minipage}
    \caption{Evolution of the least squares between the true value $\bold{M}(\bold{a}_i^{(j)})$ and its estimate as a function of the number of iterations with two normalizations (no normalization, normalization as in Algorithm~\ref{qks-alg}). (a): the pair $(p,r)$ is set to $(2,1)$. (b): the pair $(p,r)$ is set to $(1,2)$.}
    \label{f2}
\end{figure}

In \figref{f1}-(a) and \figref{f2}-(a), the pair $(p,r)$ is set to $(2,1)$. \figref{f1}-(a) plots the residual of the QUARKS cost function as a function of the iteration number for both normalized and non-normalized algorithms. Convergence to a global minimum is observed for both cases. The convergence towards a unique fixed point is shown with \figref{f2}-(a) which displays the least-squares residual between the true value $\bold{M}(\bold{a}_i^{(j)})$ and its estimate. When normalizing the columns of the factor matrix, the factor matrices converge to their true values while it is not the case for the non-normalized version. Although both algorithms reach a global minimum, the solution to the QUARKS identification problem is not unique as highlighted with Figure~\ref{f2}-(b), and both solutions are equivalent as they provide a similar prediction-error (up to machine precision).

The case $(p,r) = (1,2)$ is analyzed in \figref{f1}-(a) and \figref{f2}-(b). We observe in the two latter figures that using normalization affects the convergence speed to a global minimum. In this example, about 500 iterations were required: this observation is very much case dependent. However, for all experiments carried out, the non-normalized algorithm converged to a global minimum in few iterations.

The algorithm converges in a monotonous manner to a fixed point for both pairs $(p,r)$. In the next example, we illustrate the performance on a practical case study for different scenarii of normalization and regularization.
 
\subsection{Case study 2: Adaptive optics}
The wavefront aberrations are here generated according to \cite{Beghi:2011}. In this paper, two layers of turbulence with different statistics and windspeed are located on conjugated planes and added up to form the wavefront measured by the sensor. The atmospheric turbulence is a stochastic process, therefore 50 realizations are carried out. Unless mentioned otherwise in the next paragraphs, the default parameters for AO simulations are listed in Table~\label{qks-table}. 
\begin{table}[h!]
%    \captionsetup{justification=centering}
    \centering
    \begin{tabular}{l c}
    \hline
    \textbf{Turbulence} & \\ 
    Number of layers & 2 \\
    Fried parameter, $r_0$ & $\{0.2, 0.4\} \left[ m \right]$ \\    
    Outer scale, $L_0$ & $10  \left[ m \right]$ \\  
    Number of phase points per lenslet, $n_\phi$ & 3 \\
	Horizontal windspeed, & $\{1,2\} \left[ \text{points/sample} \right]$\\
	Number of realizations, & 50 \\
    \hline
    \textbf{Telescope} & \\
    Telescope aperture, $D$ & $1 \left[ m \right]$ \\
    Sampling frequency & $500 \left[ Hz \right]$ \\ 
    Number of lenslets, $N$ & 10 \\
    Number of sensor measurements, $2N^2$ & 200 \\
    Signal-to-Noise Ratio, SNR & $15 \left[ dB \right]$ \\ \hline
    \end{tabular}
    \caption{Default parameters for AO simulation.}
    \label{qks-table}
\end{table}

\subsection{Benchmark methods and quality criteria}
%Studying the noise sample covariance $E\left[\bold{e}_k\bold{e}_k^T \right]$ in \eqref{VARX} is outside the scope of this paper. 
Three methods for identification are compared:
\begin{enumerate}
\item unstructured least squares
\begin{equation}
\min_{\bold{A}_i} \sum_{k=p+1}^{N_t} \| \bold{s}(k) -  \sum_{i = 1}^p \bold{A}_i \bold{s}(k-i) \|_2^2 
\end{equation}
\item regularized sparse least-squares using \cite{Boyd:sparse}:
\begin{equation}
\label{sparse_pb}
\min_{\bold{A}_i} \sum_{k=p+1}^{N_t} \| \bold{s}(k)  -  \sum_{i = 1}^p \bold{A}_i \bold{s}(k-i) \|_2^2 + \tau \sum_{i = 1}^p \| \text{vec}(\bold{A}_i) \|_1
\end{equation}
where $\tau$ is a regularization parameter.
\item QUARKS identification \eqref{CF} with Algorithm~\ref{qks-alg} without normalization. No knowledge of the normalization coefficients is available. Algorithm~\ref{qks-alg} is initialized only once, randomly.
The stopping criterion parameters $\epsilon$ and $I_{max}$ are set respectively to $10^{-5}$ and $3$. The maximum number of iterations $\kappa_{max}$ is $100$. The hyperparameters were randomly searched within the bounds mentioned in Section~\ref{sec14} and within the range $\left[0,5\right]$ for $(\lambda,\mu)$: the set of hyperparameters over 20 realizations that yields the lowest prediction-error is selected. The curse of dimensionality that appears when choosing hyperparameters with grid search is bypassed with random search, \cite{rdm}. Bayesian optimization or online non-linear optimization for hyperparameter estimation are outside the scope of this paper. 
\end{enumerate}

The performances are checked on a validation dataset containing $5 \times 10^3$ temporal points. The results are discussed based on the Variance Accounted For (VAF) between the signals $\bold{s}(k+1)$ and $\widehat {\bold{s}}(k+1) = \sum_{i = 1}^p \widehat{\bold{A}_i} \bold{s}(k-i)$: 
\[
\text{VAF}(\bold{s}(k),\widehat{\bold{s}}(k) ) = \text{max}\bigl( 0,\bigl( 1- \frac{\frac{1}{N_t}\sum_{k = 1}^{N_t}\|\bold{s}(k) -\widehat{\bold{s}}(k)  \|_2^2  }{\frac{1}{N_t}\sum_{k = 1}^{N_t}\| \bold{s}(k) \|_2^2} \bigr) \times 100\bigr)
\]
The VAF between two identical signals $\bold{s}(k)$ and $\widehat{\bold{s}}(k) $ reaches $100 \%$. 
The experiments are carried out on MatlabR2016b using a desktop computer with a CPU Intel Xeon E5-2609V2/2.5 GHz.

In Subsection~\ref{qks:exp1}, we compare three structures for large-scale VARX modeling and analyze the impact of increasing the spatial order $r$ for QUARKS identification. In Subsection~\ref{qks:exp2}, we investigate the impact of temporal and spatial regularization for varying SNR conditions. We illustrate the computational complexity analysis with timing experiments for a range of network sizes in Subsection~\ref{qks:exp3}. 

\subsection{Illustration of QUARKS identification}
\label{qks:exp1}

The identification set contains $5 \times 10^3$ temporal measurements. The temporal order of the VAR model is set to $2$. We first choose a Kronecker rank within $r = \{1,...,5\}$. The parameters $\lambda$ and $\mu$ in \eqref{CF} are set to 0. The minimization \eqref{sparse_pb} is solved for $\tau = \text{logspace}(0,4,8)$. 

\begin{figure}[h!]
%\captionsetup{justification=centering}
		\centering
        \includegraphics[width=0.85\linewidth]{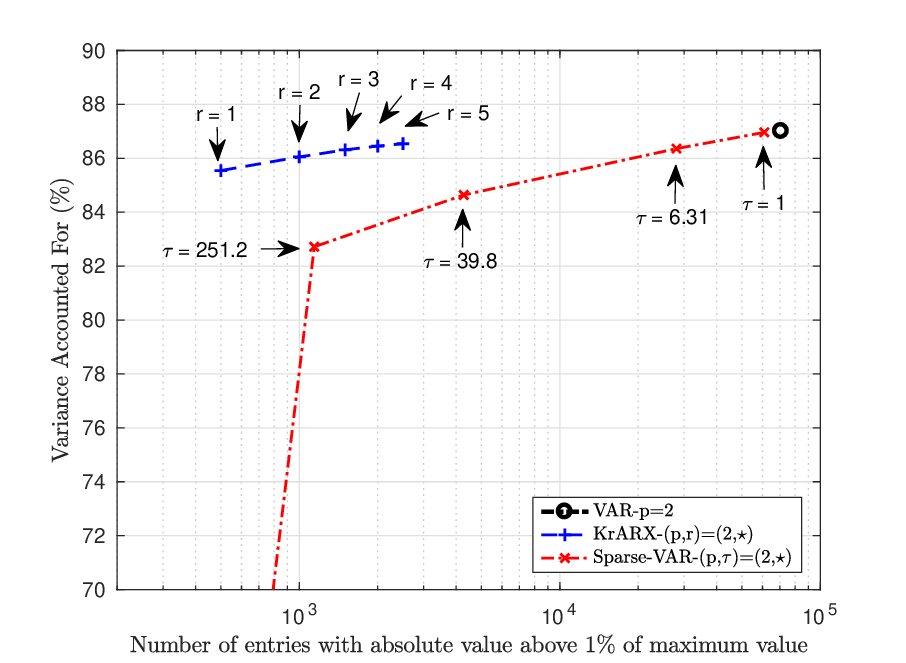}
        \caption{Variance Accounted For ($\%$) versus complexity of model. A blue cross corresponds to an estimate with given Kronecker rank. Each red cross corresponds to a regularization parameter $\tau$ on the sparsity prior in \eqref{sparse_pb}. Two points are not visible on the plot: $(\tau,\% \text{non-zero values},\text{VAF}) \in \{(1.5849\times 10^3,389,45.3), (10^4,124,0) \}$.}
\label{tradeoff}
\end{figure}

We define a measure that we call \emph{model complexity} as the number of non-zero entries needed to construct the $p$ coefficient matrices. For example, the complexity of a QUARKS model is at most $2prN^2$ (only the non-zero elements of the factor matrices) while it reaches a total of $pN^4$ for the full least squares estimation. It is illustrated in Figure \ref{tradeoff} that displays the VAF with respect to the number of non-zero elements (with truncated entries at $1\%$ of the maximum value) needed to construct the full coefficient matrix $\bold{A}_1$. The prediction error is computed on a validation dataset \emph{after} truncation. 

We emphasize that no truncation on the elements of the factor matrices is done for the Kronecker model. For example, a total of $500$ non-zero values are necessary to build the Kronecker factors associated to $\bold{A}_1$ and reaches $85.54\%$ accuracy. 
The VAF obtained with the sparse identification decreases with increasing regularization parameter $\tau$ as expected while the number of non-zero entries decreases for a high prior on sparsity. 

This trade-off between the complexity of the model and the accuracy of the prediction error is present in the QUARKS modeling as well. While the estimated matrix with $\ell_1$ minimization tries to reduce the number of non-zero entries, the matrix obtained with QUARKS modeling does not exhibit sparse patterns but a prominent multi-level structure. The lower the spatial order $r$, the lower the model complexity and the higher the prediction error is. 

\subsection{Influence of the hyperparameters}
\label{qks:exp2}

The regularization with $r_s$ and $r_t$ in \eqref{QUARKScost_vec} is the most beneficial with short data batches or in noisy environments. The difference with the case $(\lambda,\mu) = (0,0)$ is all the more significant when the ratio $\frac{N_t}{Nrp}$ is approximately 1. 
The parameters for this subsection are gathered in Table~\ref{qks:tbl}. 
\begin{table}[h!]
%\captionsetup{justification=centering}
    \begin{center}
    \begin{tabularx}{0.45\textwidth}{l c}
    \hline
    \textbf{Sensor} & \\
    Signal-to-Noise Ratio, SNR $\left[ dB \right]$ & $\{ 5, 10, 15, 20, 25, 30, 35, 40 \}$ \\ \hline
    \textbf{QUARKS} & \\
    Temporal order, $p$ & 4 \\
    Spatial order, $r$ & 2 \\
    Number of points for identification, $N_t$ & 500 \\  \hline
    Number of Monte-Carlo simulations & 50 \\ \hline
    %\textbf{Regularization} & \\
    %Weight for temporal stability, $\mu$ & $\{0, \text{logspace}(-6,-3,5)\}$ \\
    %Decay, $\eta$ & $\{0.5,0.6,0.7 \}$\\
    %Smoothness, $\xi$ & $\{0.5,0.6,0.7 \}$ \\
    %Weight for spatial stability, $\lambda$ & $\mu $\\
    %Decay, $\zeta$ & $\text{linspace}(10^{-3},10^{-1},3)$ \\ \hline
    \end{tabularx}
        \caption{Parameters for identifying QUARKS model in \ref{qks:exp2}.}
        \label{qks:tbl}
    \end{center}
\end{table}

\begin{figure}[h!]
		\centering
        \includegraphics[width=0.85\linewidth]{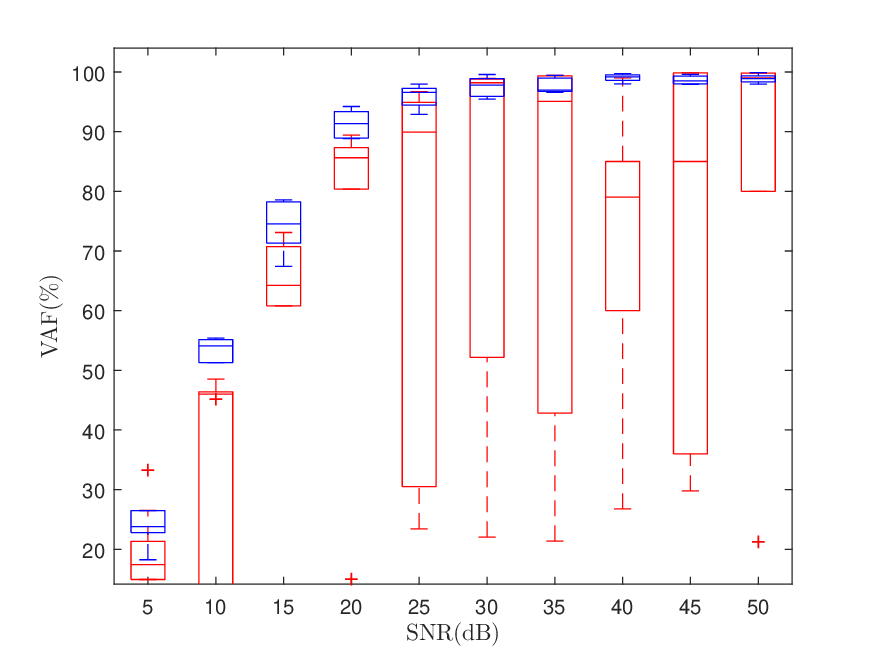}
        \caption{Variance Accounted For ($\%$) versus the signal-to-noise ratio. Red: without regularization nor normalization. Blue: with both regularization and normalization to 1.}
        \label{regplot}
\end{figure}

\figref{regplot} displays the VAF on validation data with and without regularization. Regularizing the cost function in noisy situations and with relatively few data samples leads to substantial improvements over the non-regularized QUARKS identification. It especially reduces the variance of the prediction error while the performance of the non-regularized version with few temporal samples is very unreliable. Random search has interesting performances as it exploits the fact that some hyperparameters may not contribute a lot for obtaining good solutions in the example at hand.  

\subsection{Scalability}
\label{qks:exp3}

One advantage of the new modeling paradigm is to reduce the computational complexity for estimating large-scale VARX models. No regularization is considered in this section in order to analyze whether the QUARKS identification in Algorithm~\ref{qks-alg} scales with $\mathcal{O}(N^3N_t)$.  

\begin{figure}[h!]
		\centering
        \includegraphics[width=0.85\linewidth]{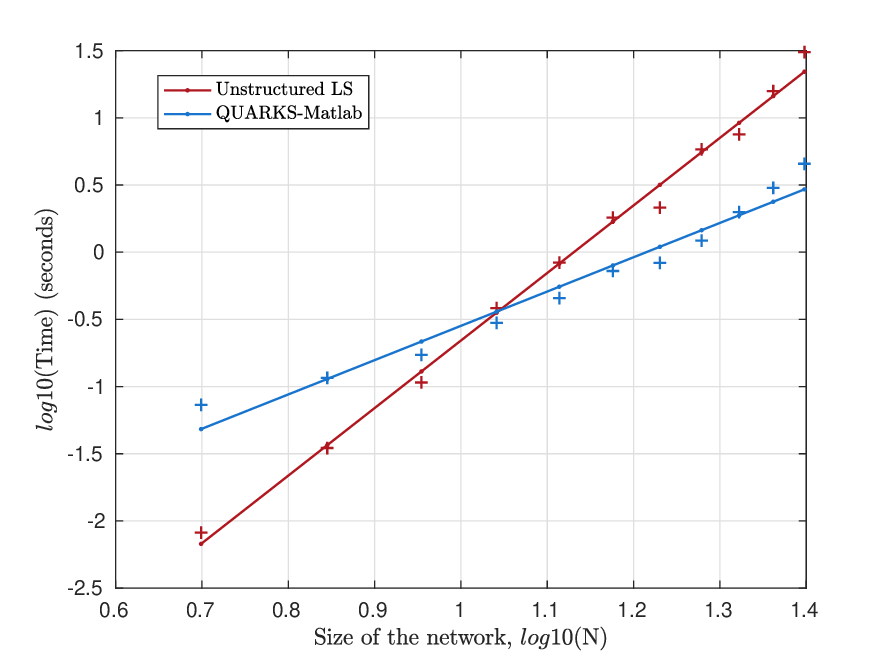}
        \caption{Evolution of the computational time with respect to the size of the 2D array. The linear model fitted with the QUARKS method is: $log10(\text{Time}) = 2.55 \times log10(N) -3.10 $, $\sigma = 0.34$ while it is: $log10(\text{Time}) = 5.03 \times log10(N) -5.68 $, $\sigma = 0.27$ with the unstructured least-squares (LS).}
        \label{scalafig}
\end{figure}

The temporal order is set to $4$ and Kronecker rank to $2$. The number of lenslets $N$ belongs to the range $\left[5:2:29\right]$, which implies $2 \times \{5^2,...,29^2\}$ sensor signals at each time sample. The number of time samples for QUARKS identification is such that $N_t = 10 p r N$ while it is $N_t = 50N^2$. These values were fixed such that the prediction-error is similar for both methods. The linear model fitted in \figref{scalafig} for the QUARKS has a regression coefficient of $3.27$ (with standard deviation $\sigma = 0.51$) while the unstructured estimation has a coefficient of $5.18$ ($\sigma = 0.50$). 

Although the QUARKS implementation includes many loops and would take advantage of a C implementation, the reduction in the regression coefficient is significant using the Kronecker-based identification.

\section{Conclusions}
In this paper, the class of Kronecker networks is defined and the VAR modeling part is investigated. Each coefficient matrix of the VAR model is approximated with a sum of few Kronecker matrices which offers high data compression for large networks. 
Estimating in least-squares sense the data matrices give rise to a bilinear problem which is addressed using Alternating Least Squares. The convergence of ALS to a fixed point was proven in very particular conditions and assuming persistency of excitation and non-zero initial guesses. Further structure on the factor matrices can be added. Numerical examples on atmospheric turbulence prediction, e.g for AO control, demonstrates the high compression capabilities of this model as well as its scalability for larger networks.

The algorithm has been presented for 2D dynamical systems and can easily be generalized to higher dimensions by using a Kronecker product of multiple matrices instead of only two matrices in which case larger compression rates are achieved. Such higher order modeling for 2D arrays is obtained by tensorizing the sensor data $\bold{S}(k)$ and allows to establish a new trade-off between accuracy and computational complexity. 

\begin{appendices}

\section*{Appendix. Proof for Theorem~\ref{thm-qks}}
In this appendix, we derive the proof of convergence for the regularized ALS with a very particular normalization. The proof builds on \cite{Li:15} and therefore, we only highlight the main changes here compared to the vector form.
\newline 
\emph{Notations.} The noise term $\bold{\overset{\sim}{E}}$ is defined similarly as $\bold{\overset{\sim}{S}}$ from the noise components $\bold{e}(k)$. Moreover, $\bold{\overset{\sim} U} = \bold{\overset{\sim} U}_1, \quad \bold{\overset{\sim}{s}} = \text{vec}(\bold{\overset{\sim}{S}}), \quad \bold{M} = \bold{I}_N \otimes \bold{\overset{\sim} U}$. The iteration counter $\kappa$ is left out.
\newline 
\newline 
\noindent
First, an inner product for matrices in $\mathbb{R}^{(N(N_t-1)+N^3) \times N}$ is defined.
\begin{defn} \label{app-def4}
Let $\bold{X},\bold{Y} \in \mathbb{R}^{N \times N}$ and denote their columns with $\bold{x}_i,\bold{y}_i$.
For two matrices $\overline{\bold{X}}, \overline{\bold{Y}}$ such that: 
\[
\overline{\bold{X}} = 
\bold{M}
\begin{bmatrix}
\bold{I}_N \otimes \bold{x}_1  \\ \vdots \\ \bold{I}_N \otimes \bold{x}_N
\end{bmatrix} 
\]
and similarly for $\overline{\bold{Y}}$, the inner product on $\mathbb{R}^{(N(N_t-1)+N^3) \times N}$ is defined with:
\[
\langle \overline{\bold{X}} , \overline{\bold{Y}} \rangle = \lambda_{max}(\bold{\overset{\sim} U}^T\bold{\overset{\sim} U} ) {\rm vec}(\bold{X})^T{\rm vec}(\bold{Y}) 
\]
\end{defn}
\begin{lem} \label{app-ip} For the matrix $\overline{\bold{X}} \in \mathbb{R}^{(N(N_t-1)+N^3) \times N}$ and the inner product in Definition~\ref{app-def4}, the quantity $\| \overline{\bold{X}} \|_2 = \sqrt{\langle \overline{\bold{X}} , \overline{\bold{X}} \rangle}$ is a norm on $\mathbb{R}^{(N(N_t-1)+N^3) \times N}$. 
\end{lem}
\noindent
\textbf{Proof.}  The proof contains four points.
\begin{enumerate}
\item $\| \overline{\bold{X}} \|_2$ is positive because the spectral radius and the Euclidean norm are both positive.
\item If $\| \overline{\bold{X}} \|_2 = 0$, and with $\bold{\overset{\sim} U}^T\bold{\overset{\sim} U}\neq 0$, then $\| \bold{x} \|_2 = 0$ and $\bold{x} = 0$. This implies that  $\overline{\bold{X}} = 0$.
\item Let $\alpha \in \mathbb{R}$. $\| \alpha \overline{\bold{X}} \|_2^2 = \lambda_{max}\bigl(\alpha^2(\bold{\overset{\sim} U}^T\bold{\overset{\sim} U})\bigr) \| \bold{x} \|_2^2 = | \alpha | \| \bold{X} \|_2^2$
\item The triangular inequality reads:
\begin{eqnarray*}
\|\overline{\bold{X}} + \overline{\bold{Y}} \|_2 &=& \sqrt{ \lambda_{max}(\bold{\overset{\sim} U}^T\bold{\overset{\sim} U})} \| \bold{x}+\bold{y} \|_2 \\
& \leq &  \| \overline{\bold{X}} \|_2 + \|\overline{ \bold{Y}} \|_2
\end{eqnarray*} 
using the triangular inequality on the Euclidean norm. \qed
\end{enumerate} 
For example, the matrix $\bold{\overline{M}_b}$ has the structure of $\overline{\bold{X}}$ in Definition~\ref{app-def4}. 

We define two sets associated to the true values $\Bell$ and $\Br$:
\begin{eqnarray*}
X_{\Bell} &=& \{ \widehat \Bell \in \mathbb{R}^{N^2}  | \forall i \in \{1,...,N\}, \| \widehat{\bold{a}_i} \|_2 \leq \| \bold{a}_i \|_2  \} \\
X_{\Br} &=& \{ \widehat{\Br} \in \mathbb{R}^{N^2} |  \forall i \in \{1,...,N\}, \| \widehat{\bold{b}_i} \|_2 = \| \bold{b}_i \|_2,  \widehat{b}_1 > 0   \} 
\end{eqnarray*}
Let $\widehat \Bell \in X_\bold{a}, \widehat{\Br} \in X_\bold{b}$.
 
\subsection*{Endomorphism}
We now prove the operator $\mathcal{F}$ maps $X_{\Bell}$ to $X_{\Bell}$. The solution of one least-squares update is written with:
\begin{eqnarray}
\widehat \Bell &=& \mathcal{F}_3(\widehat{\bold{b}})  \nonumber \\
&=& \bigl(\bold{I}_{N} \otimes (\bold{\overline{M}_{\widehat b}}^T\bold{\overline{M}_{\widehat b}} )^{-1}\bold{\overline{M}_{\widehat b}}^T \bigr)\bold{\overset{\sim}{s}}  \nonumber \\
\label{app-eq1}
&=& \bigl(\bold{I}_{N} \otimes (\bold{\overline{M}_{\widehat b}}^T\bold{\overline{M}_{\widehat b}} )^{-1}\bold{\overline{M}_{\widehat b}}^T \bigr)
\bigl((\bold{I}_{N} \otimes \bold{\overline{M}_{b}})\Bell+\underbrace{\text{vec}\bigl(\begin{bmatrix} \bold{\overset{\sim}{E}} \\ 0 \end{bmatrix} \bigr)}_{\bold{e}}\bigr) 
\end{eqnarray}
Using the partition of $\bold{a}$ into the $N$ vectors $\bold{a}_i$ of size $N$, we rewrite \eqref{app-eq1}:
\begin{equation}
\widehat \Bell_i =  (\bold{\overline{M}_{\widehat b}}^T\bold{\overline{M}_{\widehat b}} )^{-1}\bold{\overline{M}_{\widehat b}}^T  \bold{\overline{M}_{b}}\Bell_i + \bold{e}_i
\end{equation}
which corresponds to the vector form studied in \cite{Li:15}. We assumed the noise has a finite variance when $N_t$ goes to infinity which implies:
\begin{equation}
\lim_{N_t \rightarrow \infty} \| (\bold{\overline{M}_{\widehat b}}^T\bold{\overline{M}_{\widehat b}} )^{-1}\bold{\overline{M}_{\widehat b}}^T \|_2 \| \bold{e}_i \|_2 = 0
\end{equation}
Therefore, the Euclidean norm of $\widehat \Bell_i$ is upper-bounded as follows:
\begin{eqnarray}
\lim_{N_t \rightarrow \infty} \| \widehat \Bell_i \|_2 & \leq & \lim_{N_t \rightarrow \infty} \| (\bold{\overline{M}_{\widehat b}}^T\bold{\overline{M}_{\widehat b}} )^{-1}\bold{\overline{M}_{\widehat b}}^T \bold{\overline{M}_{b}} \|_2 \| \Bell_i \|_2  \nonumber \\
&\leq &  \lim_{N_t \rightarrow \infty}  \frac{ \| \bold{\overline{M}_{\widehat b}}^T\bold{\overline{M}_{b}} \|_2 } { \| \bold{\overline{M}_{\widehat b}}\bold{\overline{M}_{\widehat b}}\|_2 } \| \Bell_i \|_2  \nonumber \\
&\leq &  \lim_{N_t \rightarrow \infty} \frac{ \| \bold{\widehat b}^T \bold{ b}\|_2 } { \| \bold{\widehat b}^T \bold{\widehat b} \|_2 } \| \Bell_i \|_2 
\end{eqnarray}
The last inequality is obtained using the definition of the inner product in Lemma~\ref{app-ip}. We conclude with the following lemma.
\begin{lem} \label{lemA2} Let $\bold{b}, \widehat{\bold{b}} \in \mathbb{R}^{N}$. If $\| \widehat \Br \|_2 = \| \Br \|_2$, then 
$\| \widehat \Br^T \Br \|_2 \leq \| \widehat \Br^T \widehat \Br\|_2$.
The inequality is strict if $\widehat \Br \neq \epsilon \Br$ for $\epsilon \in \{-1,1\}$.
\end{lem}
$\widehat \Br \in X_{\Br}$ implies $\| \widehat \Br \|_2 = \| \Br \|_2$ and therefore, $\| \widehat \Bell_i \|_2 \leq \| \Bell_i \|_2$ when $N_t$ goes to infinity. The functional $\mathcal{F}(.)$ maps $X_{\Bell}$ to $X_{\Bell}$.

\subsection*{Upper bound on $Q$}
We now introduce the quantity $Q = \| \frac{d\mathcal{F}}{d\widehat \Bell} \|_2$. 
From $\widehat \Bell^{(\kappa+1)} = \mathcal{F}_3 (\mathcal{F}_2 ( \mathcal{F}_1 ( \widehat \Bell^{(\kappa)} ) ) )$, we decompose:
\begin{equation}
\label{deriv}
Q = \norm{  \frac{d\mathcal{F}}{d\widehat \Br}  \frac{d \widehat \Br}{d \widehat \Br_{\bold{n}}} \frac{d \widehat \Br_{\bold{n}}}{d \widehat \Bell} }_2 
 \leq  \norm{\frac{d \mathcal{F}_{3}}{d\widehat \Br}  }_2 \norm{ \frac{d \mathcal{F}_{2}}{d\widehat \Br_{\bold{n}}} }_2 \norm{ \frac{d \mathcal{F}_{1} }{d\widehat\Bell} }_2
\end{equation}
We further detail each norm in \eqref{deriv} and start the analysis with $\| \frac{d\mathcal{F}_{3}}{d\widehat \Br} \|_2$.
\begin{lem} (\cite{Li:15})Let $f(.)$ be defined with $f(\widehat \Br):= \bold{I}_{N} \otimes (\bold{\overline{M}_{\widehat b}}^T\bold{\overline{M}_{\widehat b}})^{-1}\bold{\overline{M}_{\widehat b}}^T$.
Under Assumption \textbf{A2}, the magnitude of the directional derivative of $f(\widehat \Br)$ along a vector $\bold{u}$ attains its maximum when $\bold{u}$ is in the same direction as $\widehat \Br$.
\end{lem} 
When taking the derivative of $f$ with respect to $\widehat \Br$, the maximum norm is obtained when the gradient is taken along the direction of $\widehat \Br$, i.e a deviation from $\Br$, denoted with $\Delta \bold{b}$, is in the same direction as $\widehat \Br$. Using the derivations from the previous section and introducing a normalized deviation $\overrightarrow{\Br}$ equal to $\frac{\Delta \bold{b}}{\| \Delta \bold{b} \|_2}$:
\begin{equation}
\label{app-deriv}
\norm{ \frac{d\mathcal{F}_{3}}{d\widehat \Br} }_2 
\leq  \frac{ \| \overrightarrow{\Br}^T \Br \|_2 }{ \| \widehat \Br^T \widehat \Br\|_2 } \| \Bell \|_2 
\end{equation}
From the definition of the unit vector $\overrightarrow{\Br}$, it can be expressed as a function of $\widehat \Br$ with $\| \widehat \Br^T \Br \|_2 = \| \overrightarrow{\Br}^T \Br \|_2 \| \Br \|_2$. Then, \eqref{app-deriv} is written as:
\begin{equation}
\label{app-derivf3}
\norm{ \frac{d\mathcal{F}_{3}}{d\widehat \Br} }_2 
\leq  \frac{ \|  \widehat \Br^T \Br \|_2 }{ \| \widehat \Br^T \widehat \Br\|_2 } \frac{\| \Bell \|_2 }{\| \Br \|_2}
\end{equation}
Now evaluating the derivative of $\mathcal{F}_2$ related to the normalization step, we write: 
\begin{equation}
\norm{ \frac{d \mathcal{F}_{2}}{d\widehat \Br_{\bold{n}}} }_2 = \norm{ \frac{d \widehat \Br}{d \widehat \Br_{\bold{n}}} }_2 \leq \frac{\| \Br \|_2}{\| \widehat \Br_{\bold{n}} \|_2}  
\end{equation}
We need to relate $\| \Br \|_2$ and $\| \widehat \Br_{\bold{n}} \|_2$.

\begin{lem} For all $i \in \{1,...,N\}, \| \widehat \Bell_i \|_2 = \| \Bell_i \|_2$ and $ \| \widehat{\bold{b_n}}_i \|_2 = \| \Br_i \|_2$.
\end{lem}
\noindent
\textbf{Proof.} 
Asymptotically,
\[
\label{update11}
\widehat \Bell =  \bigl(\bold{I}_{N} \otimes (\bold{\overline{M}_{\widehat b}}^T\bold{\overline{M}_{\widehat b}} )^{-1}\bold{\overline{M}_{\widehat b}}^T \bold{\overline{M}_{b}} \bigr) \Bell
\]
and therefore, for all $i\in\{1,\hdots,N\}$:
\[
\widehat \Bell_i =  (\bold{\overline{M}_{\widehat b}}^T\bold{\overline{M}_{\widehat b}} )^{-1}\bold{\overline{M}_{\widehat b}}^T \bold{\overline{M}_{b}}  \Bell_i
\]
Multiplying by $\bold{\overline{M}_{\widehat b}}$ on both left sides and using similar arguments as in \cite{Li:15}, $\bold{\overline{M}_{\widehat b}} \begin{bmatrix} \widehat \Bell_1 & \hdots & \widehat \Bell_N \end{bmatrix} = \bold{\overline{M}_{ b}} \begin{bmatrix} \Bell_1 & \hdots & \Bell_N \end{bmatrix}$. The right-hand side term reads:
\[
\bold{\overline{M}_{b}} \begin{bmatrix} \Bell_1 & \hdots & \Bell_N \end{bmatrix} = \bold{M}
 \begin{bmatrix} \bold{I}_N \otimes \bold{b}_1 \\ \vdots \\ \bold{I}_N \otimes \bold{b}_N \end{bmatrix} \begin{bmatrix} \bold{a}_1 & \hdots & \bold{a}_N \end{bmatrix}
\]
and hence, for all $i,j \in \{1,\hdots,N\}^2$:
\[
\bold{M}(\bold{I}_N \otimes \bold{b}_i)\bold{a}_j = \bold{M}(\bold{I}_N \otimes \bold{\widehat{b}}_i)\bold{\widehat{a}}_j
\]
The matrix $\bold{M}$ is full column rank, it follows:
\begin{eqnarray*}
(\bold{I}_N \otimes \bold{b}_i)\bold{a}_j &=& (\bold{I}_N \otimes \widehat{\bold{b}}_i)\widehat{\bold{a}}_j \\
 \bold{b}_i {\bold{a}_j}_k  &=&   \bold{\widehat{b}}_i \bold{\widehat{a}}_{j_k} 
\end{eqnarray*}
Therefore, since $\bold{a}_{j_k} \in \mathbb{R}$ and $\bold{b} \in X_{\Br}$, it follows:
$\| \bold{b}_i \|_2 | \bold{a}_{j_k} |  =  \| \bold{\widehat{b}}_i \|_2 | \bold{\widehat{a}}_{j_k} |$ and then, $|\bold{a}_{j_k}| = |\bold{\widehat{a}}_{j_k}|$, for all $k$ in the set $\{1,...,N\}$. Finally, it comes $\| \bold{a}_j \|_2 = \| \bold{\widehat{a}}_j \|_2$.
A similar reasoning starting from the relation between $\widehat{\bold{b_n}}$ and $\Br$ yields $\| \widehat{\bold{b_n}}_i \|_2 = \| \Br_i \|_2$.
\qed \newline 
We can conclude:
\begin{equation}
\label{app-derivf2}
\norm{ \frac{d \mathcal{F}_{2}}{d\widehat \Br_{\bold{n}}} }_2 \leq 1 
\end{equation}
Therefore, we use \eqref{app-derivf3} and \eqref{app-derivf2} to upper-bound the constant $Q$ with:
\begin{equation}
Q \leq  \frac{ \| \overrightarrow{\Br}^T \Br \|_2 }{ \| \widehat \Br^T \widehat \Br\|_2 } \| \Bell \|_2 \frac{ \| \overrightarrow{\Bell}^T \Bell \|_2 }{ \| \widehat \Bell^T \widehat \Bell\|_2 } \| \Br \|_2 
 \leq \frac{ \| \widehat \Br^T \Br \|_2 }{ \| \widehat \Br^T \widehat \Br\|_2 }\frac{ \| \widehat \Bell^T \Bell \|_2 }{ \| \widehat \Bell^T \widehat \Bell\|_2 } 
\end{equation}
We conclude that $Q < 1$ using Lemma~\ref{lemA2}.  

\end{appendices}

\end{document}